\documentclass[12pt]{article}
\pdfoutput=1
\usepackage{titling}
\newcommand{\cO}[1]{}
\usepackage{natbib}
\usepackage{xcolor,hyperref}
\usepackage{amsthm}
\usepackage{comment}
\definecolor{darkblue}{rgb}{0.0,0.0,0.7}
\hypersetup{colorlinks,breaklinks,linkcolor=darkblue,urlcolor=darkblue,anchorcolor=darkblue,citecolor=darkblue}

\usepackage{geometry}
\newtheorem{assumption}{Assumption}

\newtheorem{definition}{Definition}

\newtheorem{lemma}{Lemma}

\newtheorem{proposition}{Proposition}
\newtheorem{obs}{Observation}

\usepackage{units}
\usepackage{enumerate}
\usepackage{booktabs}
\usepackage{tabularx}
\usepackage{arydshln}
\usepackage{amssymb}
\usepackage{amsmath}
\usepackage{xcolor}
\usepackage{graphicx}
\usepackage{subfigure}
\usepackage{tikz}
\usepackage{pgfplots}
\usepackage{csquotes}
    \usepackage{caption}
\usetikzlibrary{decorations.pathreplacing,calligraphy}
\pgfplotsset{width=10cm,compat=1.9}
     \newcommand{\mul}{\underline{\mu}}
     \newcommand{\muh}{\overline{\mu}}
    \newcommand{\M}{\mathcal{M}}
    \newcommand{\thetal}{\underline{\theta}}
     \newcommand{\thetah}{\overline{\theta}}
     \newcommand{\pri}{\mu_0}
     \newcommand{\s}{S}
     \newcommand{\infl}{influence}
     \newcommand{\Infl}{Influence}
     \newcommand{\Bln}{Balanced}
     \newcommand{\bln}{balanced}
     \newcommand{\Unb}{Unbalanced}
     \newcommand{\unb}{unbalanced}
     \newcommand{\half}{\frac{1}{2}}
     \newcommand{\tstar}{\theta^*}
     \newcommand{\tstarb}{\theta_B^{**}}
     \newcommand{\tstaru}{\theta_U^{**}}
    \newcommand{\conI}{(I)}
    \newcommand{\conIR}{(IR)}
    \newcommand{\conIC}{(IC)}
    \newcommand{\conIRD}{(IR-$\downarrow$)}
    \newcommand{\conIRU}{(IR-$\uparrow$)}

\definecolor{green}{HTML}{30AE17}
\usetikzlibrary{graphs}

\title{
The Design and Price of Influence
}
\author{Raphael Boleslavsky   \and Aaron Kolb\thanks{Boleslavsky: Department of Business Economics \& Public Policy, Indiana University (e-mail: \texttt{rabole@iu.edu}); Kolb: Department of Business Economics \& Public Policy, Indiana University (e-mail: \texttt{kolba@iu.edu}). We acknowledge the paper ``The Design and Price of Information" by Bergemann, Bonatti, and Smolin for inspiring the title. We thank Nageeb Ali, Matteo Camboni, Pavel Ilino, Sergei Izmalkov, Pino Lopomo, Marzena Rostek, Mehdi Shadmehr, Ron Siegel, Ran Shorrer, Mark Whitmeyer, Kun Zhang, and seminar audiences at Bonn, Penn State, Pitt/Carnegie Mellon, Wisconsin, and Indiana University for helpful comments.  Boleslavsky thanks the Weimer Faculty Fellowship at Indiana University for financial support, and Kolb thanks the Philpott Faculty Fellowship.}}
\date{\today}

\begin{document}

\begin{titlingpage}

\maketitle

\abstract{A sender with private preferences would like to influence a receiver's action by providing information through a statistical test. The technology for information production is controlled by a monopolist intermediary, who offers a menu of tests and prices to screen the sender's  type. We characterize the intermediary's optimal screening menu and the associated distortions, which may benefit the receiver by increasing test informativeness. Because of these distortions, seemingly unfavorable changes in the prior belief may actually benefit the sender. In extensions, we study (i) a stronger intermediary who can commit to a ``coercive'' test to punish non-participation and (ii) a weaker intermediary who cannot control test design but can charge for access.}

\vspace{.3in}


\vspace{.1in}


\end{titlingpage}

\let\markeverypar\everypar
 \newtoks\everypar
 \everypar\markeverypar
 \markeverypar{\the\everypar\looseness=-2\relax}
\parskip=2.8pt

\newpage
\baselineskip=20.pt

\setcounter{page}{1}

\begin{displayquote}
The millions of people who derive pleasure and satisfaction from smoking can be reassured that every scientific means will be used to get all the facts as soon as possible.
---Tobacco Industry Research Committee, July 1, 1954
\end{displayquote}

\section{Introduction}

Vast resources are routinely spent in an attempt influence high stakes decisions. For example, when regulators must decide how to address a public health crisis, firms and organizations fund scientific research to steer policy decisions in their favor. Similarly, when lawmakers seek to promote sustainable energy by subsidizing infrastructure for renewable energy, interest groups hire think tanks and consultants to provide information about the environmental and economic impact of proposed projects. Brands also frequently conduct organized public relations and advertising campaigns to influence consumer and investor beliefs, hoping to increase sales or attract capital.

Three themes emerge in many of these complex interactions. First, the party that seeks to influence (``the sender'') often does not have the ability to  produce information that could persuade the decision maker (``receiver'') on his own. Instead, those seeking influence often rely on intermediaries that have the expertise or technology to conduct transparent research and disseminate it to the target audience. The concentration of market power over information production by such professional influencers allows them to appropriate surplus and has the potential to distort the information they offer. Second, large scale decisions are often complex, with heterogeneous effects that generate diverse preferences. While conservationists may agree that both solar and nuclear energy are effective alternatives to fossil fuels, they may nevertheless disagree about which renewable is most desirable. Those concerned with economic development might favor large scale investments associated with expansion of nuclear power, while consumer advocates or those concerned about disposal of nuclear waste might instead prefer solar. Third, the motives of the sender are often private information, even when the sender's identity is known. For example, political nonprofits and other 501(c) organizations that rely heavily on donations are not required to disclose their funding sources. Thus, it may be apparent that a 501(c) group favors a transition to renewable energy, but it may be more difficult to discern exactly which type of renewable it would like to promote.\footnote{Similarly, activist shareholders, including hedge funds, can privately acquire substantial stakes in the stocks of multiple companies. Indeed, under United States SEC law, disclosure of ownership is not required until a 5\% threshold is reached. Such overlapping positions can create a complex exposure to a regulator's, policymaker's, investor's, or consumer's decision.} In other words, the sender's preference for the receiver's action, which drives his willingness to pay for influential information, is not only diverse but also private. Therefore a professional influencer with market power faces both an information design problem when persuading the receiver, and a screening problem when extracting surplus from the sender.

To study the interplay of these themes, we introduce a three-player model in which a sender pays an intermediary to influence a receiver on his behalf. The receiver faces a choice between three alternatives: risky actions $L$ and $R$, and a safe option $\s$ (delay, outside option, status quo). The receiver's preferences depend on a binary state of the world. Risky actions $L$ and $R$ are optimal for the receiver if the belief is sufficiently low or high, respectively, but at the prior belief, $\s$ is optimal. The sender's preferences over the receiver's action do not depend on the state of the world, but they do depend on a privately known preference parameter, i.e., the sender's type.\footnote{An implication is that the receiver's decision does not depend on the receiver's belief about the sender's type. Nonetheless, because information is generated through the intermediary, and the intermediary must screen the sender's type, the sender's type and its distribution indirectly affect the information that is available to the receiver.} In particular, the sender's payoff from $\s$ is normalized to 0;  his payoffs from actions $L$ and $R$ are linear in his type, and the sum of these payoffs is constant and positive.\footnote{We do not require that \textit{both} $L$ and $R$ are always preferred over $\s$. To the contrary, we allow for sender types that have a negative payoff from $L$ or $R$ and a positive payoff from the other. We discuss variations on this specification and provide a microfoundation in Section \ref{dis:pay}.} Thus, the sender's preferences are horizontally differentiated: ``left-leaning'' sender types benefit more from action $L$ than $R$, while ``right-leaning'' types benefit more from $R$ than $L$.
 
 The intermediary can influence the receiver by conducting a flexible experiment that reveals information about the state of the world, as in the literature on Bayesian persuasion and information design \citep{KG2011, BM2016,BM2019, T2019}. The intermediary offers the sender a menu of experiments and prices, from which he selects his preferred alternative. The intermediary conducts the experiment on the sender's behalf, and the receiver observes the experiment and its outcome before choosing an action. Initially, we focus on settings where the sender's participation or consent is required for the intermediary to produce information.
 
 To use methods from mechanism design, we recast the intermediary's problem. Rather than choosing a statistical experiment or test, the intermediary offers the sender a menu of \textit{\infl{} bundles} and prices, where each bundle specifies the probabilities that the receiver will select actions $L$ and $R$. Obviously, any bundle that the intermediary offers must be consistent with the receiver's optimal decision for some underlying statistical experiment---we explicitly characterize the set of \infl{} bundles that can be generated in this manner. By doing so we can formulate a mechanism design problem which has several interesting features: a multidimensional allocation, countervailing incentives, and a non-standard feasible set.

We then derive the intermediary's revenue-maximizing menu of \infl{} bundles and prices. Under mild assumptions, the intermediary's optimal menu offers three experiments. One of these maximizes the probability that the receiver chooses action $L$, which is efficient for relatively low types. The second bundle maximizes the probability of action $R$, which is efficient for relatively high types. The third test induces actions $L$ and $R$ with equal probability and otherwise induces action $S$. Such a test is inefficient for all types of sender, but it is equally valued by all types of sender, which allows the intermediary to fully extract the surplus that it generates. While the optimal menu always has this structure, the matching between sender types and tests, as well as the design of the third test, depend significantly on the receiver's decision problem, which determines which influence bundles are feasible.

For moderate prior beliefs about the state, the environment is \textit{balanced} in the following sense: the experiment that maximizes the probability of the receiver choosing action $L$ also results in action $L$ more often than action $R$ (and vice versa). In this case the sender's willingness to pay for his efficient test is non-monotone in his type, first decreasing, then increasing. In other words, senders with moderate types have the lowest willingness to pay, creating an incentive to mimic toward the interior of the type space. The optimal menu offers the efficient bundles to senders with sufficiently low or sufficiently high types, but rather than exclude the moderate types, the intermediary offers such types a test that induces actions $L$ and $R$ with probability one half and never induces action $S$. All sender types have the same value for this bundle, which is fully extracted by the intermediary. Given the  optimal menu's structure, the probability that each action is implemented is monotone in the sender's type---a  type that benefits more from a particular action is (weakly) more likely to induce it.

For lower prior beliefs about the state, the environment is \textit{unbalanced}: even the bundle that maximizes the probability of action $R$ is more likely to result in $L$ than $R$. In this setting, the sender's willingness to pay for his first best test is monotone decreasing in type. In the optimal menu, the participation constraint binds for an interval of the highest types. Such types are offered a test that results in $L$ and $R$ with equal probability, but unlike in the balanced case, the probability of action $\s$ is also positive; thus, randomization over three posterior beliefs is required. From an information design perspective, this is unusual, since each sender type's ideal test is supported on two posterior beliefs and induces action $\s$ with probability zero. In other words, the need to screen produces a strong distortion of the test offered to an interval of high types. Furthermore, the optimal menu also produces an intriguing pattern of distortions in the assignment of tests to types.\footnote{This assignment is determined both by the design of the tests in the optimal menu and by their prices.} In particular, an interval of types whose preferences lean toward $L$ nevertheless purchase the experiment that maximizes the probability of action $R$, resulting in a non-assortative matching between sender types and experiments.\footnote{We are unaware of existing work in mechanism design where some types choose a different type's efficient allocation, even though their own efficient allocation is also being offered.} By implication, the probability that the receiver selects action $R$ is non-monotone in the sender's type: high types, who have the most to gain from action $R$, are less likely to convince the receiver to select $R$ than moderate types. More broadly, the optimal menu features an alternating pattern of efficiency, then inefficiency, then efficiency, and again inefficiency as type increases. 

The comparative statics also highlight intriguing differences between the balanced and unbalanced cases. As the prior belief shifts in favor of action $L$, it becomes more difficult to persuade the receiver to select $R$ and less difficult to persuade the receiver to select $L$. If the sender controlled information provision, left-leaning types  would benefit from such a shift, while right-leaning types would lose. In the balanced case, the optimal screening menu inherits this feature. However, in the unbalanced case, such a shift in the prior generates a Pareto improvement for all sender types. Furthermore, under a mild assumption on the type distribution, an interval of right-leaning types benefits as the prior shifts against action $R$. This surprising finding is a consequence of the structure of the optimal screening menu, in which the participation constraints bind for an interval of the highest types.

To determine the intermediary's impact on the receiver's welfare, we study a benchmark with unmediated communication, where information provision maximizes the sender's payoff. This environment might arise from a change in disclosure rules that make sender's type transparent to the intermediary, or changes to market structure (for example, a merger or increased competition) that give sender greater authority over information provision. We show that such structural changes may reduce the receiver's welfare. In other words, the distortions in information provision that the intermediary introduces to screen may increase the receiver's value of information.

In the rest of the paper, we examine two model variations that expand or contract the set of instruments available to the intermediary. In the first extension, we allow the intermediary to \textit{coerce} the sender by committing to produce information (conduct a specific experiment) if the sender does not participate in the mechanism. Effectively, the intermediary designs an outside option whose value to the sender depends on his type. Such coercion has no benefit for the intermediary if all sender types weakly prefer both $L$ and $R$ to $\s$ (the receiver's default action). But if some senders obtain disutility from, say, $L$, then there is a potential benefit to designing an outside option that induces $L$ with some probability. Such coercion is costly, however, because it can undermine the incentive for other types (who obtain positive utility from $L$) to pay for experiments that increase the probability of action $L$, since they obtain some influence for free. In light of this cost, the intermediary's coercive outside option is distorted relative to a world in which the sender's type is known. We provide sufficient conditions for coercion to be beneficial for the intermediary and characterize the optimal coercive menu. We further show that such coercion can increase information quality and the receiver's welfare.

Our second extension moves the opposite direction, weakening the intermediary by removing her ability to design information. In this extension, authority over test design is held by the sender, but the intermediary can nevertheless charge the sender for access to the receiver. We discuss the channels by which the inability to design information affects the intermediary's screening problem, and we show that it can degrade information quality and reduce the receiver's welfare.

\paragraph{Related Literature.} Broadly, our paper lies at the intersection of two fundamental lines of research: information design \citep{KG2011,BM2016,BM2019,T2019} and monopolistic screening  \citep{mussa1978monopoly,myerson1981optimal}. Relative to information design, our main novelty is transferring control of information production from the sender to the intermediary, generating a screening problem for the latter. This screening problem exhibits a number of non-standard features including a multidimensional allocation, countervailing incentives, and an endogenous feasible set, which must be characterized as part of the analysis.

Within this strand of literature, the two most closely related papers are \citet{BBS2018} and \citet{ali2022sell}. \citet{BBS2018} study a monopoly screening problem in which the decision maker buys information directly from a data broker, who must screen the decision maker's prior belief about the state. In that paper, interior types have the most uncertainty about the state and thus the highest willingness to pay for information. The optimal menu always offers a fully informative signal to a subset of these types. Under conditions identified by the authors, the optimal menu may also include a less informative (distorted) signal intended for extreme types. A fundamental difference in our problem is that our information-buyer is distinct from the decision maker, and a fully informative signal is generically not part of an optimal menu. Furthermore, in our model, either (i) extreme types want to mimic moderate types or (ii) all types want to mimic in the same direction, so distortions in the optimal menu follow a different pattern.\footnote{If our sender had private information about the state of the world, as in \citet{BBS2018}, a potentially interesting complication would arise: the receiver would learn not only from the results of the experiment but the choice of the experiment itself, since the latter reveals the element of a partition in which the sender lies. This partitioning problem itself is already nontrivial without experiments, even when restricting to interval partitions; see recent work by \citet{onuchic2023conveying}.} In \citet{ali2022sell}, an agent who holds an asset can purchase information about its quality from an intermediary. Before selling the asset, the agent can choose whether to disclose or withhold the information, and the market cannot distinguish whether information was withheld or not acquired. The intermediary designs a signal structure, a testing fee, and a disclosure fee to maximize his revenue. Crucially, the agent does not have private information ex ante, so the intermediary does not need to screen. However, the private information that the agent acquires leads to a strategic disclosure problem that the intermediary must anticipate when designing and pricing information.

Our analysis also relates to a literature that studies mechanisms for selling multiple products (bundling). While part of this literature considers a multi-dimensional type, such problems are typically challenging \citep{A1996,RC1998,RS2003}. Other papers, including \citet{DM1996} and \citet{LMB2011} study the sale of multiple products, with preferences characterized by one-dimensional types. \citet{BIL2021} study a multiproduct screening problem where the buyer's type represents a location on the Hotelling line, characterizing the optimal sales mechanism with linear, concave, and convex transportation costs. \citet{LM2024} analyze a related model, which incorporates competition between buyers. Interpreting each component of the influence bundle (the probabilities of $L$ and $R$) as a separate product, our problem could be interpreted as one of multiproduct screening. Our setting nevertheless exhibits a crucial difference: the intermediary cannot independently vary the allocation of each product. Indeed, the components of the influence bundle are derived from an information design in the receiver's decision problem, and they must satisfy constraints which have no analogue in product design.

Our study of coercion connects to a literature on auctions with externalities, in which the seller threatens to allocate the item to a rival if a buyer does not participate \citep{jehiel1996not,jehiel1999multidimensional,figueroa2009role}. These threats rely on externalities among buyers: with a single buyer and free disposal, it is difficult for a seller of an item to affect the buyer's outside option. In contrast, our intermediary sells influential information; the sender cannot free himself of the consequences of the receiver's action, even if he chooses not to buy. Provided the intermediary can produce information without the sender's consent, she may be able to commit to a coercive test, which will be run in the event of non-participation.

Finally, a number of papers also combine information design and contracting, with a different focus. Rather than surplus extraction by the party that designs information, these papers focus on incentivizing an agent to produce or reveal information in a manner desired by the principal. \citet{HT2022} and \citet{Y2022} study a contract offered by a principal, who hires an agent to produce information that is relevant to her decision problem. In \citet{HT2022} the agent's effort cost is known, but he can secretly distort the experiment and lie about its outcome.\footnote{See also \citet{WZ2022}, wherein a sender privately produces flexible information, and can then pay an intermediary to certify it.} Conversely, in \citet{Y2022}, the experiment is observable and its outcome is hard information, but the agent can misrepresent his cost of effort. In a different vein, \citet{DPS2023} study a principal who would like to elicit, garble, and then transmit an informed agent's private information to an uninformed receiver. The principal's communication with the agent must be public, and she must therefore incentivize the agent to garble his announcement in the way that she would like.

\section{Model}
We study a game with three players:  sender (he), intermediary (she), and receiver (it).  There is a binary state of the world $\omega\in\Omega\equiv\{0,1\}$. None of the three players knows $\omega$, and they share a common prior $\pri\equiv\Pr(\omega=1)$.

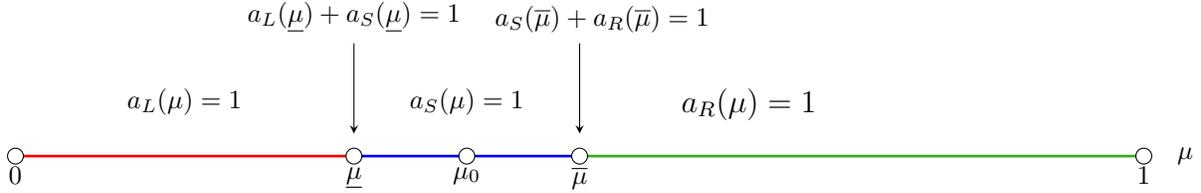
\begin{figure}
\begin{center}
\begin{tikzpicture}[scale=1.5]
\draw[line width=1pt, color=red] (0,0) -- (3,0); 
\draw[line width=1pt, color=green] (10,0) -- (5,0); 
\draw[line width=1pt, color=blue] (3,0) -- (5,0); 

 \draw [stealth-](3,0.2) -- (3,1);
 \fill (3,1) node[above] {\footnotesize{$a_L(\mul)+a_\s(\mul)=1$}};

 \draw [stealth-](5,0.2) -- (5,1);
 \fill (5.2,1) node[above] {\footnotesize{$a_\s(\muh)+a_R(\muh)=1$}};

\fill (0,0) node[below] {\footnotesize{$0$}};
\fill (10,0) node[below] {\footnotesize{$1$}};
\fill (5,0) node[below] {\footnotesize{$\muh$}};
\fill (10.2,0) node[right] {\footnotesize{$\mu$}};
\fill (3,0) node[below] {\footnotesize{$\mul$}};
\fill (4,0) node[below] {\footnotesize{$\pri$}};

\filldraw[color=black, fill=white] (4,0) circle (2pt);
 
 \filldraw[color=black, fill=white] (0,0) circle (2pt);
 \filldraw[color=black, fill=white] (10,0) circle (2pt);
\filldraw[color=black, fill=white] (5,0) circle (2pt);

\filldraw[color=black, fill=white] (3,0) circle (2pt);

  \fill (6.5,0.7) node[below] {$a_R(\mu)=1$};
 
   \fill (1.5,0.7) node[below] {\footnotesize{$a_L(\mu)=1$}};
 

  \fill (4,0.7) node[below] {\footnotesize{$a_\s(\mu)=1$}};
\end{tikzpicture}
\caption{Receiver's Decision Problem}\label{Fig:RD}
\end{center}
\end{figure}

\paragraph{Receiver.} The receiver chooses a single action $a\in A\equiv\{\s,L,R\}$. By selecting $\s$ (safe), the receiver obtains a payoff of 0 regardless of the state. Actions $L$ and $R$ deliver a cost or benefit to the receiver depending on whether her action matches the state, 
\begin{align*}
v(S,\omega)&=0\\
    v(L,\omega)&=(1-\omega)\mul-\omega(1-\mul)=\mul-\omega\\
    v(R,\omega)&=-(1-\omega)\muh+\omega(1-\muh)=\omega-\muh,
\end{align*}
where $0\leq \mul<\pri<\muh\leq 1$. Let $\mu\equiv\Pr(\omega=1)$ be the receiver's belief and $a_i(\mu)$ for $i\in A$ be the optimal probability of action $i$ at belief $\mu$. The receiver's optimal strategy is summarized in Figure \ref{Fig:RD}. To simplify the exposition, we have presented a particular payoff function $v(\cdot,\cdot)$ for the receiver. Our results, including those that deal with receiver welfare, rely only on the structure of the receiver's optimal strategy, not on the specific parametrization of $v(\cdot,\cdot)$. Provided the receiver's optimal strategy is maintained, $v(\cdot,\cdot)$ can vary with no changes to the results.

\paragraph{Sender.} The sender has a stake in the receiver's decision, represented by quasi-linear preferences $u(a,\theta)-p$, where
\begin{align*}
    u(\s,\theta)=0\quad\quad u(L,\theta)=1-\theta\quad\quad u(R,\theta)=\theta.
\end{align*}
Preference parameter $\theta$ is sender's private information, drawn from a set $\Theta\subset \mathbb{R}$ according to cumulative distribution function $F$ with continuous density $f>0$.  A \textit{moderate} sender has $\theta\in[0,1]$, and an \textit{extremist}, $\theta>1$ or $\theta < 0$. Thus, a moderate sender weakly prefers both $L$ and $R$ over $\s$, while an extremist prefers only one of them. A sender is \textit{left-leaning} if he prefers $L$ to $R$, i.e.,  $\theta<\half$, \textit{right-leaning} if  $\theta>\half$, and \textit{neutral} if $\theta=\half$.

Since all sender types prefer the receiver to select some action other than its default action under the prior, $\s$, they all benefit from the opportunity to influence the receiver's action by conducting an informative experiment or test that reveals information about the state of the world, $\omega$. A sender's \textit{willingness to pay} for any such experiment, however, generally varies with his type.

\paragraph{Intermediary.} The testing technology is controlled by an intermediary, who can attempt to persuade the receiver on the sender's behalf in exchange for payment. We represent tests or statistical experiments using the \textit{obedience approach}, which is commonly encountered in the literature on information design \citep{BM2016, T2019}. This approach exploits an equivalence between statistical experiments and \textit{obedient decision rules} $\Pi\,:\Omega\,\rightarrow\Delta(A)$. Such rules recommend an action to the receiver according to a distribution that varies with the state, with the property that it is incentive compatible for the receiver to follow the recommendation.\footnote{Consider a statistical test $\pi\,:\Omega\rightarrow \Delta(X)$, which assigns a probability distributions over some outcome space to each state of the world. The obedience approach is built on the insight that regardless of the outcome of the experiment, it is the receiver's action that determines the players' payoffs. 
Thus, if multiple test outcomes induce the receiver to select the same action, these can be replaced by a single outcome that reports the receiver's preferred action. With this replacement, following the recommended action (obedience) is incentive compatible for the receiver. See \citet{BM2016,BM2019,T2019} for more information.} Given the equivalence established in the literature, we refer to obedient decision rules as ``tests'' or ``experiments.''

Because the intermediary does not know the sender's type, she offers a screening menu $\mathcal{M}=\{\Pi(\hat{\theta}),p(\hat{\theta})\}_{\hat{\theta}\in \Theta}$ consisting of a test $\Pi(\hat{\theta})$ and a price $p(\hat{\theta})$ for each reported type $\hat{\theta}\in\Theta$. Note that the test's price does not depend on its outcome---this is without loss since the state and test outcomes are independent of the sender's true type.\footnote{Because sender is risk neutral and all players have a common prior over the state, the intermediary could always replace any outcome-dependent prices with their ex ante expectation without violating any constraints. If sender types were correlated with the state, then the intermediary could benefit from outcome-dependent prices that encode ``side bets" as in \cite{cremer1988full}. However, outcome-dependent prices could give the intermediary an incentive to manipulate the test's findings in order to increase revenue. Such an incentive might cause the sender and receiver to question the credibility of the test, undermining the intermediary's ability to extract surplus.} Our goal is to characterize the intermediary's revenue-maximizing screening menu.

\paragraph{Timing.} The timing of the game is as follows
\begin{enumerate}
    \item[(i)] The intermediary posts a menu $\mathcal{M}$. 
    \item[(ii)] The sender privately learns his type $\theta$. 
    \item[(iii)] The sender observes the mechanism and chooses whether to participate. If he does not, then no new information is released, i.e., the intermediary's test is uninformative.
    \item[(iv)] If the sender participates, then he issues a report $\hat{\theta}$, thereby selecting test $\Pi(\hat{\theta})$ and paying $p(\hat{\theta})$.
    \item[(v)] Receiver observes the test and the realized recommendation and selects an action.  
\end{enumerate}
Note that in the model above, if sender chooses not to participate, then no new information is (or can be) generated. In Section \ref{sec:coercion} we extend the mechanism by allowing the intermediary to \textit{coerce} the sender: the intermediary designs a test $\Pi(n)$ that she would conduct if the  sender chooses not to participate.

\paragraph{Alternative Interpretation.} Motivated by our leading examples, we have presented the model as monopolistic screening in the market for influential information. However, the model has other interpretations, where the sender's payment is not literally a monetary transfer. For instance, the sender could be an intern, the intermediary a manager, and the receiver a future employer. The sender has private horizontal preferences over job placement and enters into an agreement to supply a particular level of costly effort (the payment) for the manager in exchange for a recommendation tailored to the sender's preferences.

\section{Preliminaries}\label{sec:pre}

In this section we formulate the intermediary's optimization as a mechanism design problem.

\paragraph{From Tests to \Infl{} Bundles.} Because the sender's preferences depend only on the receiver's action, it is convenient to work with the ex ante probability distribution over actions induced by a test. Composing the prior belief $\pri\in\Delta(\Omega)$ with the test $\Pi:\Omega\,\rightarrow\Delta(A)$ generates an unconditional probability distribution over actions, $q\in\Delta(A)$. We refer to the last two components of $q$, the ordered pair $(q_L,q_R)$ as the \textit{\infl{} bundle} that is \textit{implemented} or \textit{induced} by the test. Thus, for any given test, $q_i$ is the probability that the receiver selects action $i\in\{L,R\}$. Of course, the requirement that obedience is incentive compatible restricts the set of \infl{} bundles that can be implemented. We refer to the set of \infl{} bundles that can implemented as the \textit{implementable set}, denote it $Q$, and characterize it explicitly in Lemma \ref{lem:feasible_q_pairs}.

\paragraph{Intermediary's Problem.} To formulate the intermediary's problem, let 
\begin{align*}
    u(q_L,q_R,\theta)\equiv (1-\theta)q_L+\theta q_R
\end{align*}
represent the type-$\theta$ sender's willingness to pay for a test which induces \infl{} bundle $(q_L,q_R)$. The intermediary's problem is to design a menu 
\begin{align*}
\mathcal{M}=\{(q_L(\theta),q_R(\theta),p(\theta))\}_{\theta\in\Theta}
\end{align*}
to solve
\begin{align*}
   \max_{\mathcal{M}} \int_{\Theta} p(\theta) \,dF(\theta)
\end{align*}
subject to implementation, incentive compatibility, and individual rationality constraints: for all $\theta\in \Theta,$
\begin{align}
&(q_L(\theta),q_R(\theta))\in Q\label{eq:I}\tag{I-$\theta$} 
\\
&u(q_L(\theta),q_R(\theta),\theta)-p(\theta)\geq  u(q_L(\theta'),q_R(\theta'),\theta)-p(\theta') & 
 \text{for all }  \theta'\in \Theta 
\label{eq:IC}\tag{IC-$\theta$}\\
     &u(q_L(\theta),q_R(\theta),\theta)-p(\theta) \geq 0. \label{eq:IR} \tag{IR-$\theta$}
\end{align}
The implementation constraint \eqref{eq:I} ensures that the \infl{} bundle offered to type $\theta$ can be implemented by some test. The incentive compatibility constraint \eqref{eq:IC} ensures that each sender type is willing to reveal himself truthfully. The individual rationality constraint \eqref{eq:IR} ensures type $\theta$'s participation. In an extension (see Section \ref{sec:coercion}), we allow the intermediary to coerce the sender by committing to perform test $(q_L(n),q_R(n))$ if sender chooses not to participate. In that setting, the right hand side of all \eqref{eq:IR} constraints becomes $u(q_L(n),q_R(n),\theta)$. Note that the type inside the equation label refers to the particular type $\theta$ to whom the constraint applies. When we refer to the complete set of constraints we omit it.

\paragraph{Implementable Set.}
We characterize the set $Q$ of \infl{} bundles $(q_L,q_R)$ that can be implemented by some test.

\begin{figure}
\begin{minipage}{0.5\textwidth}
\begin{tikzpicture}[scale=0.7]

\draw[line width=0.5pt] (0,0) -- (0,10); 
\draw[line width=0.5pt] (0,0) -- (10,0); 
\fill (10,0) node[right] {\footnotesize{$q_L$}};
\fill (0,10) node[above] {\footnotesize{$q_R$}};

\draw[line width=0.5pt,dotted] (0,0) -- (10,10); 
\draw[line width=0.5pt,dotted] (10,0) -- (0,10); 

\draw[line width=0.5pt,blue] (0,0) -- (0,4.16)--(4.375,5.625)--(6.875,3.125)--(5.833,0)--(0,0); 

 \filldraw[color=blue, fill=white] (0,0) circle (3pt);
\filldraw[color=blue, fill=white] (0,4.16) circle (3pt);
\filldraw[color=blue, fill=white] (4.375,5.625) circle (3pt);
\filldraw[color=blue, fill=white] (6.875,3.125) circle (3pt);
\filldraw[color=blue, fill=white] (5.833,0) circle (3pt);

\fill (0,0) node[above right,xshift=.7em] {\footnotesize{$45^\circ$}};

\fill (0,0) node[below] {\footnotesize{$(0,0)$}};
\fill (0,4.16) node[left] {\footnotesize{$R^*_0$}};
\fill (4.38,5.625) node[above] {\footnotesize{$R^*$}};
\fill (7.2,3.13) node[above] {\footnotesize{$L^*$}};
\fill (5.833,0) node[below] {\footnotesize{$L_0^*$}};

\fill (2,5.7) node[below] {\footnotesize{$(\ref{eq:feas3})$}};
\fill (6,4.5) node[above] {\footnotesize{$(\ref{eq:feas1})$}};
\fill (7,1) node[above] {\footnotesize{$(\ref{eq:feas2})$}};

\fill[blue,opacity=0.05] (0,0) -- (0,4.16)--(4.375,5.625)--(6.875,3.125)--(5.833,0)--(0,0);

\node[rotate=20] at (2,4.5) {\footnotesize{slope $\kappa_R$}};
\node[rotate=315] at (5.5,4) {\footnotesize{slope $-1$}};
\node[rotate=70] at (6,1.5) {\footnotesize{slope $\kappa_L$}};


\end{tikzpicture}
\end{minipage}
 \begin{minipage}{0.5\textwidth}
\begin{tikzpicture}[scale=0.7]
\draw[line width=0.5pt] (0,0) -- (0,10); 
\draw[line width=0.5pt] (0,0) -- (10,0); 
\fill (10,0) node[right] {\footnotesize{$q_L$}};
\fill (0,10) node[above] {\footnotesize{$q_R$}};

\draw[line width=0.5pt,dotted] (0,0) -- (10,10); 
\draw[line width=0.5pt,dotted] (10,0) -- (0,10); 

\node[rotate=20] at (4.5,2.8) {\footnotesize{slope $\kappa_R$}};
\node[rotate=315] at (7,2.5) {\footnotesize{slope $-1$}};
\node[rotate=70] at (8,0.8) {\tiny{slope $\kappa_L$}};

\draw[line width=0.5pt,blue] (0,0) -- (0,1.66)--(6.25,3.75)--(8.75,1.25)--(8.33,0)--(0,0); 

 \filldraw[color=blue, fill=white] (0,0) circle (3pt);
\filldraw[color=blue, fill=white] (0,1.66) circle (3pt);
\filldraw[color=blue, fill=white] (6.25,3.75) circle (3pt);
\filldraw[color=blue, fill=white] (8.75,1.25) circle (3pt);
\filldraw[color=blue, fill=white] (8.33,0) circle (3pt);

\fill (0,0) node[above right,xshift=.7em] {\footnotesize{$45^\circ$}};

\fill (0,0) node[below] {\footnotesize{$(0,0)$}};
\fill (0,1.66)  node[left] {\footnotesize{$R^*_0$}};
\fill (6.25,3.75) node[above] {\footnotesize{$R^*$}};
\fill (8.95,1.25) node[above] {\footnotesize{$L^*$}};
\fill (8.33,0) node[below] {\footnotesize{$L_0^*$}};

\fill (1,3) node[below] {\footnotesize{$(\ref{eq:feas3})$}};
\fill (8,2.5) node[above] {\footnotesize{$(\ref{eq:feas1})$}};
\fill (9,0) node[above] {\footnotesize{$(\ref{eq:feas2})$}};

\fill[blue,opacity=0.05] (0,0) -- (0,1.66)--(6.25,3.75)--(8.75,1.25)--(8.33,0)--(0,0);


\end{tikzpicture}
\end{minipage}
\caption{\label{Fig:Imp}Implementable Set.}
\end{figure}
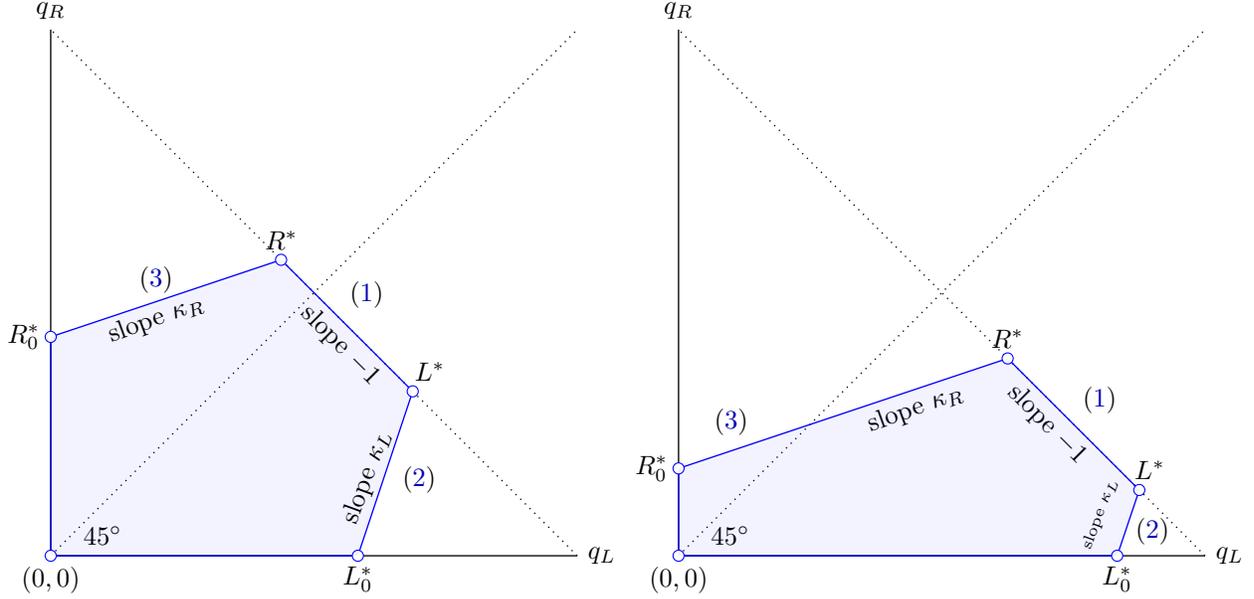

\begin{lemma}[Implementable \Infl{} Bundles]\label{lem:feasible_q_pairs}
    A pair $(q_L,q_R)\in [0,1]^2$ can be implemented by some test if and only if the following constraints are satisfied:
\begin{align}
    q_L+q_R  &\leq 1\label{eq:feas1}\\
    q_L (\muh-\mul)-q_R(1-\muh)&\leq \muh-\pri\label{eq:feas2}\\
    -\mul q_L+(\muh-\mul)q_R&\leq \pri-\mul\label{eq:feas3}.
\end{align}
\end{lemma}
Thus, we have shown that the implementable set is
\begin{align*}
Q=\{(q_L,q_R)\,|\, q_L\geq 0,\,q_R\geq 0, (\ref{eq:feas1}),(\ref{eq:feas2}),(\ref{eq:feas3})\}.
\end{align*}
Two examples of the implementable set are illustrated in Figure \ref{Fig:Imp}. We refer to the faces of the implementable set by the inequality \eqref{eq:feas1}-\eqref{eq:feas3} that binds along it. It is helpful to introduce notation for the slopes of these faces,
\begin{align}\label{slopes}
    \kappa_R\equiv \frac{\mul}{\muh-\mul}\quad\text{and}\quad\kappa_L\equiv \frac{\muh-\mul}{1-\muh},
\end{align}
where $\kappa_R$ is the slope of (\ref{eq:feas3}) and $\kappa_L$ of (\ref{eq:feas2}).

To understand the structure of the implementable set, it is helpful to consider its extreme points. Obviously, extreme point $(0,0)$ corresponds to tests that are insufficiently informative to push the receiver away from the default action $\s$. Meanwhile, extreme points 
\begin{align*}
    R^*\equiv\left(1-\frac{\pri}{\muh},\frac{\pri}{\muh}\right)\quad\text{and}\quad L^*\equiv \left(\frac{1-\pri}{1-\mul},\frac{\pri-\mul}{1-\mul}\right) 
\end{align*}
maximize the probability of actions $R$ and $L$, respectively. As described in the leading example of \citet{KG2011}, $R^*$ is induced by a test which either reveals state $0$ or  leaves the receiver with belief $\muh$, recommending $R$ in this case (and vice versa for $L^*$). Thus, a test which maximizes the probability of action $R$ ($L$) also induces $L$ ($R$) with complementary probability.\footnote{At play here is the obedience constraint: maximizing the probability of action $R$ requires that recommendation $R$ is always transmitted in state $1$. Upon observing any other recommendation, the receiver infers that the state is $0$ and would only obey recommendation $L$.} As illustrated in the right panel of Figure \ref{Fig:Imp}, it is also possible that $R^*$, despite maximizing $q_R$, is nonetheless more likely to induce action $L$ than $R$. In other words, $R^*$ can lie below the 45\textdegree{} line. (Conversely, $L^*$ can lie above it.) We will see later that the orientation of $R^*$ and $L^*$ with respect to the 45\textdegree{} line significantly affects the intermediary's optimal screening menu.  In light of this observation, we introduce the following definition.

\begin{definition}[\Bln{} and \Unb{} Environments] The environment is \textbf{\bln{}} if $R^*$ lies above the 45\textdegree{} line and $L^*$ lies below it, i.e. $\muh<2\pri<1+\mul$. The environment is \textbf{\unb{}} if both $R^*$ and $L^*$ lie below the 45\textdegree{} line, inducing a higher probability of action $L$ than action $R$, i.e., $2\pri<\muh$.\footnote{A setting in which both $L^*$ and $R^*$ lie above the 45\textdegree{} line is identical to the \unb{} environment, up to relabeling. Furthermore, it is clearly not possible for $R^*$ to be below and $L^*$ above.}
\end{definition}

The remaining extreme points, 
\begin{align*}
 R_0^*\equiv \left(0,\frac{\pri-\mul}{\muh-\mul}\right) \quad\text{and}\quad  L_0^*\equiv \left(\frac{\muh-\pri}{\muh-\mul},0\right)  
\end{align*}
maximize the probability of actions $R$ and $L$, respectively, while ensuring that the receiver only selects the targeted action or $\s$, never the opposite action. For both tests, incentive compatibility binds. The posterior beliefs are supported on $\{\mul,\muh\}$ with the same  distribution, but $R^*_0$ recommends $R$ at $\muh$ and $S$ at $\mul$, while $L^*_0$ recommends $S$ at $\muh$ and $L$ at $\mul$.

Naturally, due to the linearity of the sender's preferences over influence bundles, these extreme points will play a role in the intermediary's optimal menu design problem. However, the extreme points identified thus far arise purely from the informational constraints of the problem; as we will see in the analysis that follows, the sender's incentives will impose additional constraints, giving rise to other bundles being part of the optimal screening menu.

It is also worth mentioning that the structure of the implementable set is more complex than what would arise naturally in a multi-product monopolistic screening model. Imagine a seller offering two products $L,R$ to a buyer who would like to purchase at most one of them. Allowing randomization, the feasible set of allocations is the unit simplex in $\mathbb{R}^3$, represented in Figure \ref{Fig:Imp} by the right triangle bounded by the axes and line $q_L+q_R=1$. In our setting, the implementable set  must be consistent with an underlying recommendation policy that is incentive compatible for the receiver, and therefore must satisfy additional constraints which change its structure. 

 \paragraph{Implementing Bundles via Experiments.}
Though we mainly model tests as \infl{} bundles, it is worth describing the underlying (Bayes-Plausible) distributions of posterior beliefs and incentive compatible action recommendations that induce \infl{} bundles along the boundaries of $Q$.\footnote{Unlike persuasion problems that characterize sender-optimal distributions of posteriors, we cannot assume sender-preferred tie-breaking by the receiver. Therefore, the posterior belief distribution must be coupled with an incentive compatible action recommendation when the receiver is indifferent.} Recall that bundles $L^*$ and $R^*$ are induced uniquely by splitting the receiver's beliefs into $\{\mul,1\}$ and $\{0,\muh\}$ respectively, while recommending $R$ at the higher realization and $L$ at the lower. Any \infl{} bundle along face (\ref{eq:feas1}) other than $L^*$ and $R^*$ can be represented as a convex combination of $L^*$ and $R^*$. Such a convex combination generates a distribution of posteriors supported on $\{0,\mul,\muh,1\}$, with action $L$ selected at the lower two realizations and $R$ at the higher two. However, this representation is not unique. In particular, the realizations $\{0,\mul\}$ and $\{\muh,1\}$ can be collapsed to a single realization at their centers of mass, $\{\mu_L,\mu_R\}$ respectively, without changing the probability that actions $L$ and $R$ are chosen. Furthermore, because $\mu_L\in(0,\mul)$ and $\mu_R\in(\muh,1)$, beliefs $\{\mu_L,\mu_R\}$ can be spread or contracted slightly in a way that does not change $(q_L,q_R)$, while maintaining Bayes-Plausibility. Thus, \infl{} bundles along face \eqref{eq:feas1} can be induced by many different distributions of posteriors that require only two  realizations.

For \infl{} bundles on faces \eqref{eq:feas2} and \eqref{eq:feas3}, no such multiplicity exists.  Following the same logic, any \infl{} bundle along face \eqref{eq:feas3} can be represented as a convex combination of $R^*$ and $R_0^*$. Bundle $R_0^*$ is induced uniquely by splitting the prior belief into $\{\mul,\muh\}$, with action $\s$ recommended at $\mul$ and action $R$ at $\muh$. Combining this with the unique representation of $R^*$, any convex combination of $\{R_0^*,R^*\}$ induces posteriors $\{0,\mul,\muh\}$, where $L$ is recommended at belief 0, $\s$ at $\mul$, and $R$ at $\muh$. In contrast to the preceding case, there is no scope to collapse multiple posterior realizations into one or to adjust the posterior beliefs themselves without changing the probability that an action is selected.  The optimal menu in Proposition \ref{prop:unb} will feature a test on face \eqref{eq:feas3}, which therefore requires three belief realizations.

\section{Screening}\label{sec:screening}

In this section we characterize the intermediary's optimal screening menu. We assume senders are drawn from an interval $[\thetal,\thetah]$ with cumulative distribution function $F$ and continuous probability density function $f>0$. The following assumptions streamline the exposition.

\begin{assumption}[Monotone Virtual Types]\label{assu:monotone_virtual}
    Both $\phi^+(\theta)\equiv\theta-\frac{1-F(\theta)}{f(\theta)}$ and $\phi^-(\theta)\equiv\theta+\frac{F(\theta)}{f(\theta)}$ are strictly increasing.
\end{assumption}

\begin{assumption}[Type Bounds]\label{a:bounds}   The type space contains the neutral sender: $\thetal<\half<\thetah$. In addition, the most extreme types are bounded: (1) if $\kappa_L>1$, then $\thetal>-\frac{1}{\kappa_L-1}$, and (2) if $\kappa_R<1$, then $\thetah<\frac{1}{1-\kappa_R}$.
\end{assumption}

Assumption \ref{assu:monotone_virtual} avoids complications due to ironing.\footnote{This assumption also distinguishes our findings with those of \citet{BBS2018}. In particular, with Assumption \ref{assu:monotone_virtual} and a (weakly) increasing density $f(\cdot)$, the corresponding  virtual surplus in \citet{BBS2018} (which they denote $\phi(\theta,q)$; see p. 24) is strictly increasing. In this case, the optimal menu of \citet{BBS2018} contains a single fully informative experiment (see their Corollary 1).} Assumption \ref{a:bounds} bounds the type space to simplify the first best and optimal screening menus.  Unless stated to the contrary, we maintain these assumptions throughout.

\subsection{First Best} As a benchmark, consider the first best. The intermediary maximizes the payoff of each sender type, which she fully extracts via the payment. Thus, the intermediary chooses \infl{} bundles within the implementable set to maximize 
\begin{align*}
p(\theta)=u(q_L(\theta),q_R(\theta),\theta)=(1-\theta)q_L(\theta)+\theta q_R(\theta).
\end{align*}

For moderate types, the solution is straightforward---offer $R^*$ to right-leaning moderates ($\half\leq \theta\leq 1)$ and $L^*$ to left-leaning moderates ($0\leq\theta\leq \half$). However, a sufficiently extreme right-leaning sender ($\theta>1$ sufficiently large) might dislike action $L$ so much that he would be willing to give up a substantial probability of action $R$ in order to reduce the probability of action $L$ to 0. Thus, for sufficiently extreme sender types, the first best test can be $R_0^*$ or $L_0^*$. The  bounds on the type space (A\ref{a:bounds}) ensure that the first best test for all senders is either $R^*$ or $L^*$. 

\begin{figure}
\begin{minipage}{0.5\textwidth}
\begin{tikzpicture}[scale=0.7]

\draw[line width=0.5pt] (0,0) -- (0,10); 
\draw[line width=0.5pt] (0,0) -- (10,0); 
\fill (10,0) node[right] {\footnotesize{$\theta$}};

\draw[line width=0.5pt,blue] (10,5.625) -- (5,5) -- (0,6.875); 

\fill (10,5.625) node[above] {\footnotesize{$(1,\frac{\pri}{\muh})$}};
\filldraw[color=blue, fill=white] (10,5.625) circle (3pt);

\filldraw[color=blue, fill=white] (0,6.875) circle (3pt);
\fill (0,7.1) node[right] {\footnotesize{$(0,\frac{1-\pri}{1-\mul})$}};

\filldraw[color=blue, fill=white] (5,5) circle (3pt);
\fill (5,5) node[above] {\footnotesize{$(\frac{1}{2},\frac{1}{2})$}};

\draw[line width=0.5pt, dotted] (5,5) -- (5,0); 

\fill (2.5,0) node[above] {\footnotesize{$L^*$}};
\fill (7.5,0) node[above] {\footnotesize{$R^*$}};



\end{tikzpicture}
\end{minipage}
 \begin{minipage}{0.5\textwidth}
\begin{tikzpicture}[scale=0.7]

\draw[line width=0.5pt] (0,0) -- (0,10); 
\draw[line width=0.5pt] (0,0) -- (10,0); 
\fill (10,0) node[right] {\footnotesize{$\theta$}};

\draw[line width=0.5pt,blue] (10,3.75) -- (5,5) -- (0,8.75); 

\fill (10,3.75) node[above] {\footnotesize{$(1,\frac{\pri}{\muh})$}};
\filldraw[color=blue, fill=white] (10,3.75) circle (3pt);

\filldraw[color=blue, fill=white] (0,8.75) circle (3pt);
\fill (0,9.1) node[right] {\footnotesize{$(0,\frac{1-\pri}{1-\mul})$}};

\filldraw[color=blue, fill=white] (5,5) circle (3pt);
\fill (5.5,5) node[above] {\footnotesize{$(\frac{1}{2},\frac{1}{2})$}};

\draw[line width=0.5pt, dotted] (5,5) -- (5,0); 

\fill (2.5,0) node[above] {\footnotesize{$L^*$}};
\fill (7.5,0) node[above] {\footnotesize{$R^*$}};


\end{tikzpicture}
\end{minipage}
\caption{\label{Fig:WTP}Willingness to pay for the first best test.}
\end{figure}
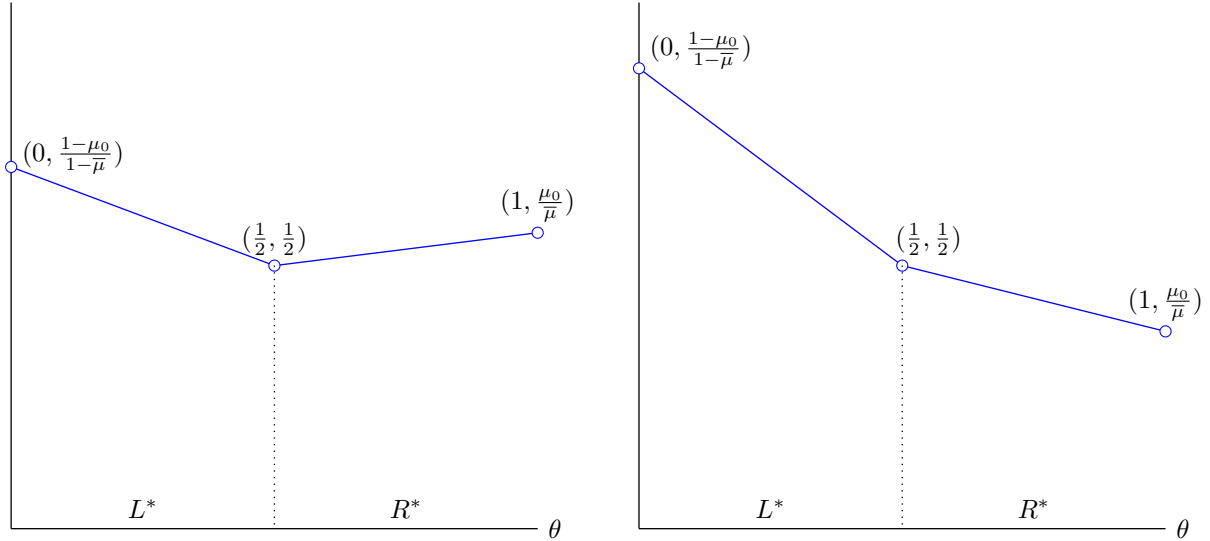

\begin{obs}[First Best]\label{obs:FB} For all $\theta\in\Theta$ the first best menu has $p(\theta)=u(q_L(\theta),q_R(\theta),\theta)$, 
\begin{align*}
(q_L(\theta),q_R(\theta))=\begin{cases}
  L^*  &\text{if}\qquad \theta<\frac{1}{2}\\
  R^* &\text{if}\qquad \theta>\frac{1}{2}.
\end{cases}
\end{align*}
\end{obs}

A fundamental feature of our environment is that a sender's willingness to pay for a particular test may be either increasing or decreasing in his type. In particular, fix a test $(q_L,q_R)$ and types $\theta>\theta'$. Note that
\begin{align*}
u(q_L,q_R,\theta)-u(q_L,q_R,\theta')=(\theta-\theta')(q_R-q_L).
\end{align*}
Thus, willingness to pay increases in type for tests which are more likely to induce action $R$ ($q_R>q_L$) and decreases in type for tests which are more likely to induce action $L$ ($q_L>q_R$).  In other words, our environment exhibits countervailing incentives \citep{LS1989}: when offered an influence bundle above the 45\textdegree{} line the sender wants to mimic down, and when below, up.

Building on this observation, consider a sender's willingness to pay for his type's first best test. In \bln{} environments (where $R^*$ and $L^*$ lie on opposite sides of the 45\textdegree{} line), willingness to pay for the first best test is V-shaped---it is lowest at type $\half$ and increases away from it, as illustrated in the left panel of Figure \ref{Fig:WTP}. By implication, in the screening menu we would expect individual rationality to bind at an interior type and local incentive compatibility to bind in the direction of this interior type. In contrast, in \unb{} environments, willingness to pay for the first best is decreasing for all types (assuming (A\ref{a:bounds})), as illustrated in the right panel of Figure \ref{Fig:WTP}. Here we expect the individual rationality constraint to bind at the highest type $\bar{\theta}$ and upward local incentive compatibility to bind for all other types.

 \subsection{Reformulating the Intermediary's Problem} Consider a menu that satisfies \conI{} and \conIR{}. Let $U(\theta)\equiv u(q_L(\theta),q_R(\theta),\theta)-p(\theta)$ denote type $\theta$'s expected utility from truthfully reporting. Standard arguments \citep{myerson1981optimal} can be used to establish that the menu satisfies \conIC{} if and only if the following conditions hold:
 \begin{align}
   q_R(\cdot)-q_L(\cdot) \quad \text{is weakly increasing}  \label{eq:MON}\tag{MON}\\
   U(\theta)=U(\theta')+\int_{\theta'}^\theta (q_R(t)-q_L(t))dt,\label{eq:INT}\tag{INT}
 \end{align}
 for all $\theta\in\Theta$, where $\theta'$ is an arbitrarily chosen reference type. In our problem, $q_R-q_L$ (the derivative of indirect utility with respect to type) plays the role of the probability of winning the item in a standard auction setup.
 
By substituting  $p(\theta)=u(q_L(\theta),q_R(\theta),\theta)-U(\theta)$ and then \eqref{eq:INT} into the intermediary's objective and simplifying in the standard way, we can rewrite the intermediary's problem as
 \begin{align}
     \max_{(q_L(\cdot),q_R(\cdot))} &\int_{\thetal} ^{\theta'}f(\theta)u(q_L(\theta),q_R(\theta),\phi^-(\theta))d\theta+\int_{\theta'} ^{\thetah}f(\theta)u(q_L(\theta),q_R(\theta),\phi^+(\theta))d\theta-U(\theta')\label{eq:objective}\\
     &\text{subject to \conI{}, \conIR{}, and \eqref{eq:MON}.} \nonumber
 \end{align}
 
We intend to relax the intermediary's problem to allow for pointwise maximization of the integrand in (\ref{eq:objective}). To do so, we must address the two constraints \conIR{} and \eqref{eq:MON}, which link bundles offered to different types.\footnote{After substituting out $p(\theta)$, (IR-$\theta$) reduces to $U(\theta)\geq 0$, which involves both $\theta$ and $\theta'$ via \eqref{eq:INT}.} In a standard monopoly screening problem \citep{mussa1978monopoly,myerson1981optimal}, such a  relaxation is achieved by dropping the monotonicity constraint on the allocation, and by replacing the participation constraints for all types with a single participation constraint for the worst-off type. In standard settings the argument is straightforward: incentive compatibility and feasibility alone imply that the buyer's indirect utility is a (weakly) increasing, (weakly) convex function of his type. Thus, the buyer with the lowest type must be worst-off, and his IR constraint is necessary and sufficient for the IR constraints of all types. Rewriting the objective using the lowest type as reference allows for pointwise maximization, and additional conditions on primitives (e.g., monotone virtual types) guarantee that the solution of the relaxed problem has a monotone allocation.

In our setting, this approach presents a number of challenges. The first challenge is to identify a type $\theta_0$ at which the individual rationality constraint binds. Indeed, if $\theta_0$ is misidentified, then imposing (IR-$\theta_0$) would not be a relaxation of the original problem, since the optimal menu of the original problem does not satisfy this constraint. Furthermore, \eqref{eq:IR} can bind at interior type(s): unlike in the standard screening problem, in our problem the constraints imply only that the sender's indirect utility is convex, not that it is increasing. The second challenge is that pointwise maximization of the integrand need not produce a menu that satisfies (\ref{eq:MON}), even when virtual types $\phi^-(\cdot)$ and $\phi^+(\cdot)$ are monotone. This issue arises because the direction of local incentive compatibility reverses at type $\theta_0$, and thus the relevant virtual type jumps down, from $\phi^-(\cdot)$ to $\phi^+(\cdot)$. Consequently, the pointwise maximizer $q_R(\cdot)-q_L(\cdot)$ may also jump down at $\theta_0$. The third challenge is related: pointwise maximization may violate the individual rationality constraint for types near $\theta_0$. For example, the pointwise maximizer may have $q_R(\cdot)<q_L(\cdot)$ for an interval of types just above $\theta_0$. In this case the sender's indirect utility is decreasing just to the right of $\theta_0$, violating the individual rationality constraint.

We address these complications with the following result.

\begin{lemma}[Relaxation]\label{lem:theta_0} (1) Let $\theta_0\equiv \half$ if the environment is \bln{} and let $\theta_0\equiv\thetah$ if the environment is \unb{}. If a menu solves the intermediary's problem, then $U(\theta_0)=0$ ((IR-$\theta_0$) binds).
(2) A menu that that satisfies (IR) and \eqref{eq:MON} satisfies the following conditions for all $\theta$
\begin{align}
    &q_R(\theta)\leq q_L(\theta)\text{ if }\theta<\theta_0&\label{eq:IR-theta_down}\tag{IR-$\theta\downarrow$}\\
    &q_R(\theta)\geq q_L(\theta)\text{ if }\theta>\theta_0. &\label{eq:IR-theta_up}\tag{IR-$\theta\uparrow$}
\end{align}
(3) Any menu that satisfies (IR-$\theta_0$), along with \eqref{eq:IR-theta_down} and \eqref{eq:IR-theta_up} for all types, also satisfies \eqref{eq:IR} for all types.
\end{lemma}

Building on Lemma \ref{lem:theta_0}, we formulate a relaxed problem in which \eqref{eq:MON} is dropped and \conIR{} is replaced by (IR-$\theta_0$),  coupled with \eqref{eq:IR-theta_down}, and \eqref{eq:IR-theta_up} for all types (denoted \conIRD{} and \conIRU{}). Writing the sender's payoff with reference to the worst-off type $\theta_0$ identified in Lemma \ref{lem:theta_0}, we formulate the relaxed problem,
\begin{align}
     \max_{(q_L(\cdot),q_R(\cdot))} &\int_{\thetal} ^{\theta_0}f(\theta)u(q_L(\theta),q_R(\theta),\phi^-(\theta))d\theta+\int_{\theta_0} ^{\thetah}f(\theta)u(q_L(\theta),q_R(\theta),\phi^+(\theta))d\theta-U(\theta_0)\label{eq:objective_relaxed}\tag{$\mathcal{R}$}\\
     &\text{subject to \conI{}, (IR-$\theta_0$), \conIRD{}, and \conIRU{}}. \nonumber
 \end{align}
As desired, the relaxed problem \eqref{eq:objective_relaxed} eliminates all constraints that link menus offered to different types. Thus, one maximizes the objective function pointwise (at each $\theta$), subject to (I-$\theta$), (IR-$\theta_0$), \eqref{eq:IR-theta_down}, and \eqref{eq:IR-theta_up}. Furthermore, points (1) and (2) of Lemma \ref{lem:theta_0} guarantee that the optimal menu in the original problem satisfies all of the constraints of (\ref{eq:objective_relaxed}). In other words, (\ref{eq:objective_relaxed}) is a true relaxation of (\ref{eq:objective}), delivering a weakly higher indirect payoff. Point (3) ensures that any menu satisfying the constraints in (\ref{eq:objective_relaxed}) automatically satisfies \conIR{}. If the optimal menu in (\ref{eq:objective_relaxed}) also satisfies \eqref{eq:MON} then the optimal menu in the relaxed problem also solves the original problem.  In the next sections, identify the optimal screening menu by solving \eqref{eq:objective_relaxed} via pointwise maximization of the integrand---in the proofs, we verify that it satisfies \eqref{eq:MON}.

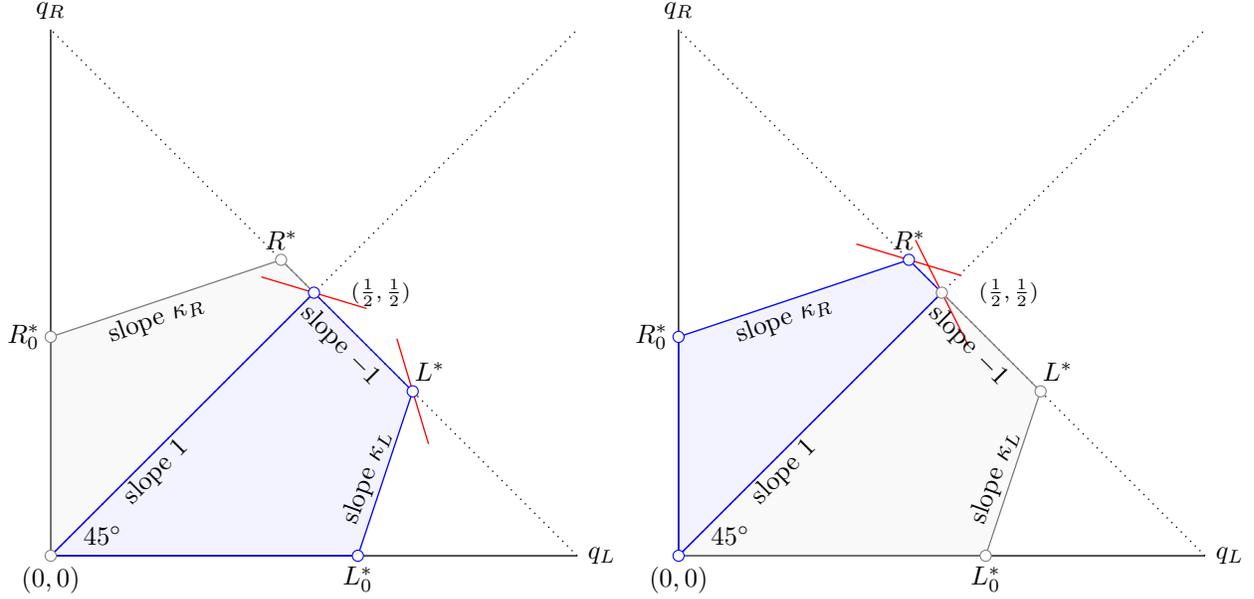
\begin{figure}
\begin{minipage}{0.5\textwidth}
    \begin{tikzpicture}[scale=0.7]

\draw[line width=0.5pt] (0,0) -- (0,10); 
\draw[line width=0.5pt] (0,0) -- (10,0); 
\fill (10,0) node[right] {\footnotesize{$q_L$}};
\fill (0,10) node[above] {\footnotesize{$q_R$}};

\draw[line width=0.5pt,dotted] (0,0) -- (10,10); 
\draw[line width=0.5pt,dotted] (10,0) -- (0,10); 

\draw[line width=0.5pt,gray] (0,0) -- (0,4.16)--(4.375,5.625)--(5,5)--(0,0); 
\draw[line width=0.5pt,blue] (0,0) -- (5,5)--(6.875,3.125)--(5.833,0)--(0,0);

\draw[line width=0.5pt,red] (5-1,5+0.3)--(5,5)--(5+1,5-0.3); 
 \draw[line width=0.5pt,red] (6.875-0.3,3.125+1)--(6.875,3.125)--(6.875+0.3,3.125-1);


 \filldraw[color=gray, fill=white] (0,0) circle (3pt);
\filldraw[color=gray, fill=white] (0,4.16) circle (3pt);
\filldraw[color=gray, fill=white] (4.375,5.625) circle (3pt);
\filldraw[color=blue, fill=white] (6.875,3.125) circle (3pt);
\filldraw[color=blue, fill=white] (5.833,0) circle (3pt);
\filldraw[color=blue, fill=white] (5,5) circle (3pt);

\fill (0,0) node[above right,xshift=.7em] {\footnotesize{$45^\circ$}};

\fill (0,0) node[below] {\footnotesize{$(0,0)$}};
\fill (0,4.16) node[left] {\footnotesize{$R^*_0$}};
\fill (4.38,5.625) node[above] {\footnotesize{$R^*$}};
\fill (7.2,3.13) node[above] {\footnotesize{$L^*$}};
\fill (5.833,0) node[below] {\footnotesize{$L_0^*$}};
\fill (5.5,5) node[right] {\scriptsize{$(\half,\half)$}};


\fill[gray,opacity=0.05] (0,0) -- (0,4.16)--(4.375,5.625)--(5,5)--(0,0);

\fill[blue,opacity=0.05] (0,0) -- (5,5)--(6.875,3.125)--(5.833,0)--(0,0);

\node[rotate=20] at (2,4.5) {\footnotesize{slope $\kappa_R$}};
\node[rotate=315] at (5.5,4) {\footnotesize{slope $-1$}};
\node[rotate=70] at (6,1.5) {\footnotesize{slope $\kappa_L$}};
\node[rotate=45] at (2,1.6) {\footnotesize{slope $1$}};


\end{tikzpicture}

\end{minipage}
 \begin{minipage}{0.5\textwidth}
\begin{tikzpicture}[scale=0.7]

\draw[line width=0.5pt] (0,0) -- (0,10); 
\draw[line width=0.5pt] (0,0) -- (10,0); 
\fill (10,0) node[right] {\footnotesize{$q_L$}};
\fill (0,10) node[above] {\footnotesize{$q_R$}};

\draw[line width=0.5pt,dotted] (0,0) -- (10,10); 
\draw[line width=0.5pt,dotted] (10,0) -- (0,10); 

\draw[line width=0.5pt,gray] (0,0) -- (5,5)--(6.875,3.125)--(5.833,0)--(0,0); 

\draw[line width=0.5pt,blue] (0,0) -- (0,4.16)--(4.375,5.625)--(5,5)--(0,0);

\draw[line width=0.5pt,red] (4.375-1,5.625+0.3)--(4.375,5.625)--(4.375+1,5.625-0.3); 
\draw[line width=0.5pt,red] (5+0.5,5-1)--(5,5)--(5-0.5,5+1);

 \filldraw[color=blue, fill=white] (0,0) circle (3pt);
\filldraw[color=blue, fill=white] (0,4.16) circle (3pt);
\filldraw[color=blue, fill=white] (4.375,5.625) circle (3pt);
\filldraw[color=gray, fill=white] (6.875,3.125) circle (3pt);
\filldraw[color=gray, fill=white] (5.833,0) circle (3pt);
\filldraw[color=gray, fill=white] (5,5) circle (3pt);

\fill (0,0) node[above right,xshift=.7em] {\footnotesize{$45^\circ$}};

\fill (0,0) node[below] {\footnotesize{$(0,0)$}};
\fill (0,4.16) node[left] {\footnotesize{$R^*_0$}};
\fill (4.38,5.625) node[above] {\footnotesize{$R^*$}};
\fill (7.2,3.13) node[above] {\footnotesize{$L^*$}};
\fill (5.833,0) node[below] {\footnotesize{$L_0^*$}};
\fill (5.5,5) node[right] {\scriptsize{$(\half,\half)$}};


\fill[blue,opacity=0.05] (0,0) -- (0,4.16)--(4.375,5.625)--(5,5)--(0,0);

\fill[gray,opacity=0.05] (0,0) -- (5,5)--(6.875,3.125)--(5.833,0)--(0,0);

\node[rotate=20] at (2,4.5) {\footnotesize{slope $\kappa_R$}};
\node[rotate=315] at (5.5,4) {\footnotesize{slope $-1$}};
\node[rotate=70] at (6,1.5) {\footnotesize{slope $\kappa_L$}};
\node[rotate=45] at (2,1.6) {\footnotesize{slope $1$}};


\end{tikzpicture}\end{minipage}
\caption{\label{Fig:BLN}Solving for Optimal Screening Menu in \Bln{} Environment.}
\end{figure}

\subsection{Optimal Menu in Balanced Environments}

First, consider \bln{} environments. The left panel of Figure \ref{Fig:BLN} illustrates optimization of the lower integral in \eqref{eq:objective_relaxed}, from $\thetal$ to $\theta_0$. The shaded blue area is the feasible set for pointwise optimization of the integrand: it is the set of \infl{} bundles that belong to $Q$ (constraint (I-$\theta$)) and lie below the 45\textdegree{} line (constraint (IR-$\theta\downarrow$)). Thus, the pointwise maximization of the lower integral is a linear program,
  \begin{align}\label{lowerLP}
      \max_{q_L(\theta),q_R(\theta)} u(q_L(\theta),q_R(\theta),\phi^-(\theta))\qquad\text{subject to}
      \qquad(q_L(\theta),q_R(\theta))\in Q\cap \{q_L(\theta)\geq q_R(\theta)\}.
  \end{align}
    As illustrated in Figure \ref{Fig:BLN}, the pointwise maximizer is either $L^*$ or $(\half,\half)$, depending on the slope of the indifference curves. In particular, the maximizer is $L^*$ if the indifference curves are steeper than face (\ref{eq:feas1}) (with slope $-1$), and it is $(\half,\half)$ if the indifference curves are flatter.\footnote{In the proof, we show that the bounds on the type space  preclude solutions at $(0,0)$ and $L_0^*$.}

  A similar characterization applies for the upper integral, from $\theta_0$ to $\thetah$. In this case, the feasible set is the subset of implementable \infl{} bundles that lies above the 45\textdegree{} line, $Q\cap \{q_L(\theta)\leq q_R(\theta)\}$, and the objective function, $u(q_L(\theta),q_R(\theta),\phi^+(\theta))$, is also linear. As illustrated in the right panel of Figure \ref{Fig:BLN}, the pointwise maximizer is $(\half,\half)$ if the indifference curves are steeper than face \eqref{eq:feas1} and $R^*$ if they are flatter.

  Building on the previous logic, it is straightforward to verify that in the lower problem the maximizer switches from $L^*$ to $(\half,\half)$ at threshold type $\tstar$, and in the upper problem the maximizer switches from $(\half,\half)$ to $R^*$ at threshold type $\tstarb$, where $\{\tstar,\tstarb\}$ are defined implicitly by
\begin{align}
    &\phi^-(\tstar)=\phi^{+}(\tstarb)= \frac{1}{2}.\label{eq:cut1}
\end{align}
Given (A\ref{a:bounds}), it is immediate that $\tstar,\tstarb$ exist and are in the interior of the type space on opposite sides of the neutral sender, $\thetal<\tstar<\half<\tstarb<\thetah$.\footnote{This follows from $\phi^-(\thetal)=\thetal<\frac{1}{2}<\phi^-(\frac{1}{2})$ and similarly $\phi^+(\frac{1}{2})<\frac{1}{2}<\thetah=\phi^+(\thetah)$, where $\phi^-$ and $\phi^+$ are continuous and strictly increasing by assumption.} 

The following proposition characterizes the optimal screening menu in balanced environments.  It is concisely summarized in the left panel of Figure \ref{Fig:utility}, which shows each type's optimal influence bundle and indirect utility for a solved example.

\begin{proposition}[Screening in Balanced Environments]\label{prop:bln}
     The optimal menu is unique. It has
    \begin{align*}
(q_L(\theta),q_R(\theta))=\begin{cases}
  L^*  &\text{if}\qquad \theta\in[\thetal,\tstar)\\
  (\half,\half)  &\text{if}\qquad \theta\in(\tstar, \tstarb)\\
  R^* &\text{if}\qquad \theta\in(\tstarb,\thetah],
\end{cases}
\end{align*}
where $\{\tstar,\tstarb\}$ are the threshold types defined in \eqref{eq:cut1}. The probability of each action is monotone in type: $q_L(\cdot)$ is decreasing, and $q_R(\cdot)$ is increasing.  Constraint (IR-$\theta$) binds for an interior interval of types, $\theta\in(\tstar,\tstarb)$.
\end{proposition}

\begin{figure}
\begin{minipage}{0.5\textwidth}
\begin{tikzpicture}[scale=0.5]


\draw[line width=0.5pt] (0,0) -- (0,10) ;
\draw[line width=0.5pt] (0,0) -- (15,0); 
\fill (15,0) node[right] {\footnotesize{$\theta$}};

\draw[line width=0.5pt,blue] (15,2.5) -- (10,0) -- (2.5,0) -- (0,.833); 
\draw[line width=0.5pt,green] (2.5,5.83) -- (0,.833+5.83); 
\draw[line width=0.5pt,green] (2.5,5) -- (10,5);
\draw[line width=0.5pt,green] (15,2.5+7.5) -- (10,7.5);

\draw[line width=0.5pt,dotted] (2.5,5.83) -- (2.5,0); 
\draw[line width=0.5pt,dotted] (10,7.5) -- (10,0); 


\filldraw[color=blue, fill=white] (10,0) circle (3pt);
\node[below] at (10,0) {$1$};


\filldraw[color=blue, fill=white] (2.5,0) circle (3pt);

\node[below] at (2.5,0) {$\frac{1}{4}$};

\fill (1.25,0) node[below] {\footnotesize{$L^*$}};
\fill (6.25,0) node[below] {\footnotesize{$(\frac{1}{2},\frac{1}{2})$}};
\fill (12.5,0) node[below] {\footnotesize{$R^*$}};

\end{tikzpicture}
\end{minipage}
 \begin{minipage}{0.5\textwidth}
\begin{tikzpicture}[scale=0.5]

 \draw[line width=0.5pt] (0,0) -- (0,10); 
\draw[line width=0.5pt] (0,0) -- (15,0); 
\fill (15,0) node[right] {\footnotesize{$\theta$}};

\draw[line width=0.5pt,blue] (15,0) -- (12.5,0) -- (2.5,1) -- (0,3.167); 

\draw[line width=0.5pt,green] (2.5,7.20) -- (0,9.33); 

\draw[line width=0.5pt,green] (2.5,5.25) -- (12.5,4.25); 

\draw[line width=0.5pt,green] (12.5,3) -- (15,3);

\filldraw[color=blue, fill=white] (12.5,0) circle (3pt);
\node[below] at (12.5,0) {$\frac{5}{4}$};


\draw[line width=0.5pt, dotted] (2.5,0) -- (2.5,7.2); 
\filldraw[color=blue, fill=white] (2.5,1) circle (3pt);
\draw[line width=0.5pt, dotted] (12.5,0) -- (12.5,4.25);

\node[below] at (2.5,0) {$\frac{1}{4}$};

\fill (1.25,0) node[below] {\footnotesize{$L^*$}};
 \fill (7.5,0) node[below] {\footnotesize{$R^*$}};
\fill (13.75,0) node[below] {\footnotesize{$B$}};

\end{tikzpicture}
\end{minipage}
\caption{\label{Fig:utility}Sender's indirect utility under the optimal screening menu for balanced (left panel) and unbalanced (right panel) environments. $U(\cdot)$ in blue, $u(q_L(\theta),q_R(\theta),\theta)$ in green. Figure constructed with $\theta\sim U[0,\frac{3}{2}]$, $\mul=\frac{1}{4}$ $\muh=\frac{2}{3}$, $\pri=\frac{1}{2}$ (balanced, left), and $\pri=\frac{3}{10}$ (unbalanced, right).}
\end{figure}

The pattern of distortions in the \bln{} environment is familiar and intuitive. When the first-best tests $R^*,L^*$ lie on opposite sides of the 45\textdegree{} line,  countervailing incentives imply that when offered the first-best tests, senders with high $\theta$ would like to mimic down, while senders with low $\theta$ would like to mimic up, feigning disinterest to reduce payment. To limit these incentives to misreport, the intermediary offers an interval of middling types an informative, but ``equalizing,'' test which induces $L$ and $R$ with probability $\half$ and never induces $\s$. All types have the same willingness to pay for this test, and since the price is set to this level, no sender can benefit from a misreport that assigns him this test.\footnote{When $\pri=\frac{1}{2}$, the equalizing test can be implemented by fully revealing the state. If it is implemented in this way, then middle types are allocated the most informative test for the lowest price in the menu, despite having the lowest willingness to pay for their first-best test. This perhaps counterintuitive finding is reconciled by the fact that our information buyer is not himself the decision maker (as in \citet{BBS2018}).}

This optimal menu has an intriguing interpretation (implementation as an ``indirect mechanism''). The intermediary establishes three ``think tanks'' or ``consultants'' that offer to conduct tests in exchange for a fee. Two of these think tanks have an extreme agenda: one only conducts test $L^*$ and the other $R^*$, at prices $p_L=u(L^*,\tstar)$ and $p_R=u(R^*,\tstarb)$, respectively. The third think tank is neutral \textit{ex post}: it only conducts an ``equalizing test,'' which implements \infl{} bundle $(\half,\half)$, at price $\half$. However, even the neutral think tank has an agenda---unless the prior belief $\pri=\half$, the equalizing test increases the probability of either $L$ or $R$ relative to full information. Furthermore, in the \bln{} environment the purchase decision is assortative---low senders ($\theta<\tstar$) hire the  left think tank, middle senders ($\theta\in(\tstar,\tstarb$)) hire the neutral think tank, and high senders hire the right think tank. Accordingly, $q_L(\cdot)$ is monotone decreasing, and $q_R(\cdot)$ is monotone increasing.

\begin{figure}
\begin{center}
\begin{tikzpicture}[scale=0.7]
\draw[line width=0.5pt] (0,0) -- (0,10); 
\draw[line width=0.5pt] (0,0) -- (10,0); 
\fill (10,0) node[right] {\footnotesize{$q_L$}};
\fill (0,10) node[above] {\footnotesize{$q_R$}};

\draw[line width=0.5pt,dotted] (0,0) -- (10,10); 
\draw[line width=0.5pt,dotted] (10,0) -- (0,10); 

\draw[line width=0.5pt,gray] (0,0) -- (0,1.66)--(2.5,2.5)--(0,0); 

\draw[line width=0.5pt,blue] (0,0) -- (2.5,2.5)--(6.25,3.75)--(8.75,1.25)--(8.33,0)--(0,0); 
  
\draw[line width=0.5pt,red] (1.5,2) -- (2.5,2.5)--(3.5,3); 
\draw[line width=0.5pt,red] (5.25,3.6) -- (6.25,3.75)--(7.25,3.9); 
\draw[line width=0.5pt,red] (8.75-0.25,1.25+0.75) -- (8.75,1.25)--(8.75+0.25,1.25-.75);

 \filldraw[color=blue, fill=white] (0,0) circle (3pt);
\filldraw[color=gray, fill=white] (0,1.66) circle (3pt);
\filldraw[color=blue, fill=white] (2.5,2.5) circle (3pt);
\filldraw[color=blue, fill=white] (6.25,3.75) circle (3pt);
\filldraw[color=blue, fill=white] (8.75,1.25) circle (3pt);
\filldraw[color=blue, fill=white] (8.33,0) circle (3pt);

\fill (0,0) node[below] {\footnotesize{$(0,0)$}};
\fill (0,1.66)  node[left] {\footnotesize{$R^*_0$}};
\fill (2.5,2.5)  node[above] {\footnotesize{$B$}};
\fill (6.25,3.75) node[above] {\footnotesize{$R^*$}};
\fill (8.95,1.25) node[above] {\footnotesize{$L^*$}};
\fill (8.33,0) node[below] {\footnotesize{$L_0^*$}};


\node[rotate=45] at (1.5,1) {\footnotesize{slope $1$}};
\node[rotate=20] at (4.5,2.8) {\footnotesize{slope $\kappa_R$}};
\node[rotate=315] at (7,2.5) {\footnotesize{slope $-1$}};
\node[rotate=70] at (8,0.8) {\tiny{slope $\kappa_L$}};

\fill[blue,opacity=0.05] (0,0) -- (2.5,2.5)--(6.25,3.75)--(8.75,1.25)--(8.33,0)--(0,0);
\fill[gray,opacity=0.05] (0,0) -- (0,1.66)--(2.5,2.5)--(0,0);

\end{tikzpicture}
\caption{\label{Fig:Unb}Solving for Optimal Screening Menu in \Unb{} Environments.}
\end{center}
\end{figure}
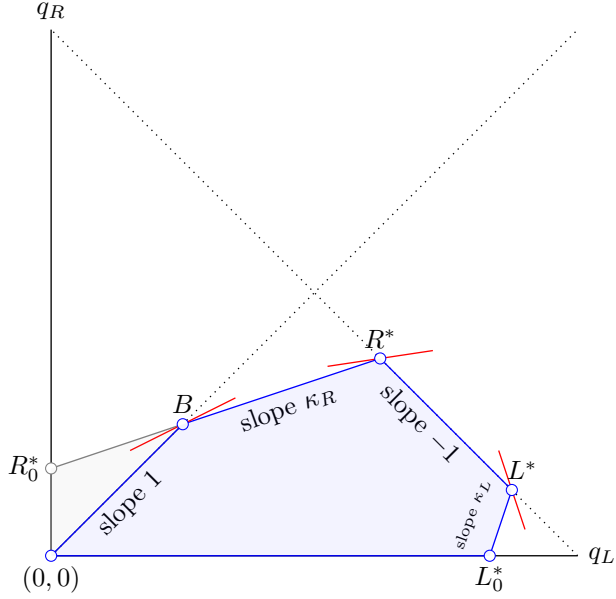

\subsection{Optimal Menu in \Unb{} Environments }
For \unb{} environments, only the left integral in \eqref{eq:objective_relaxed} is present because $\theta_0=\thetah$ (see Lemma \ref{lem:theta_0}), and the pointwise maximization is the linear program in \eqref{lowerLP}. The feasible set for the pointwise maximization is shaded blue in Figure \ref{Fig:Unb}. As in the \bln{} case, implementability (I-$\theta$) requires that a feasible bundle belongs to $Q$, while (IR-$\theta\downarrow$) requires that it is below the 45\textdegree{} line, i.e., the feasible set is $Q\cap \{q_L(\theta)\geq q_R(\theta)\}$. Unlike the \bln{} case, however, the 45\textdegree{} line cuts through $Q$ along face \eqref{eq:feas3}, rather than \eqref{eq:feas1}. In particular, the 45\textdegree{} line crosses face (\ref{eq:feas3}) at bundle $B$, where 
\begin{align*}
B\equiv \left(\frac{\pri-\mul}{\muh-2\mul},\frac{\pri-\mul}{\muh-2\mul}\right).
\end{align*}
Clearly, $B$ lies below face (\ref{eq:feas1}), and it therefore induces action $S$ with positive probability.

From our analysis of the \bln{} case, we know that for types below $\tstar$, the indifference curves are steeper than face \eqref{eq:feas1}, and the integrand is maximized at $L^*$. However, as $\phi^-(\cdot)$ increases and the indifference curves rotate counter-clockwise, the pointwise maximizer differs from that in the \bln{} case. First, for an interval of types above $\tstar$, the pointwise maximizer is $R^*$ rather than $(\half,\half)$. Second, for sufficiently large $\theta$, the indifference curves can have slope greater than $\kappa_R$, and thus the pointwise maximizer is $B$, as illustrated in Figure \ref{Fig:Unb}. Indeed, if $\phi^-(\overline\theta)\geq\frac{1}{1-\kappa_R}$, then the slope attains $\kappa_R$ at threshold type $\tstaru$ defined by
  \begin{align}
     \phi^-(\tstaru)=\frac{1}{1-\kappa_R}.\label{eq:cut3}
 \end{align}
If $\phi^-(\overline\theta) < \frac{1}{1-\kappa_R}$, then the slope never attains $\kappa_R$, and we define $\tstaru\equiv \overline\theta$ for completeness.  Note that (A\ref{a:bounds}) does not preclude an interior $\tstaru$ since $\phi^-(\theta)>\theta$ for all types above $\underline\theta$. Furthermore, for a sufficiently concentrated distribution with a symmetric, single-peaked density centered on the neutral type $\theta=\half$, it is straightforward to show that $\tstaru>\half$.\footnote{To see this, note that $\phi^-(\half)=(1/2)+F(1/2)/f(1/2)$. For a symmetric density centered on $1/2$, we have $F(1/2)=1/2$. If the distribution is sufficiently concentrated, then $f(1/2)\geq 1$ (a sufficient condition is $[\thetal,\thetah]\subseteq[0,1]$). Thus, $\phi^-(1/2)=(1/2)(1+1/f(1/2))\leq 1<1/(1-\kappa_R)$.}

The following proposition characterizes the optimal menu; the right panel  of Figure \ref{Fig:utility} illustrates a concrete example.

\begin{proposition}[Screening in Unbalanced Environments]\label{prop:unb}
 The optimal menu is unique. It has
    \begin{align*}
(q_L(\theta),q_R(\theta))=\begin{cases}
  L^*  &\text{if}\qquad \theta\in[\thetal,\tstar)\\
  R^*  &\text{if}\qquad \theta\in(\tstar, \tstaru)\\
  B &\text{if}\qquad \theta\in(\tstaru,\thetah],
\end{cases}
\end{align*}
 where $\{\tstar,\tstaru\}$ are defined in \eqref{eq:cut1} and \eqref{eq:cut3} and satisfy $\thetal<\tstar<\tstaru\leq \thetah$. The probability of action is not monotone in type: while $q_L(\cdot)$ is decreasing, $q_R(\cdot)$ is first increasing then decreasing. Constraint (IR-$\theta$) binds for high types, $\theta\in(\tstaru,\thetah)$.
\end{proposition} 

The optimal mechanism in unbalanced environments exhibits an unusual pattern of distortions: the first best  allocation is given to two disjoint intervals of types $[\thetal,\tstar)$ and $[\half,\tstaru)$, but the remaining types in the disjoint intervals $(\theta^*,\frac{1}{2})$ and $(\theta^{**}_U,\thetah]$ get distorted allocations. This pattern can be understood by considering separately the groups of left-leaning and right-leaning senders. In the latter group, the most right-leaning types---who have the lowest willingness to pay even for $R^*$---are given a distorted test $B$ to better extract revenue from moderately right-leaning types in $[\half,\tstaru)$. This is perhaps surprising, since high types have the strongest vested interest in inducing action $R$. To understand this, note that in the \unb{} environment, it is difficult to induce action $R$: the largest implementable $q_R$ is, itself, relatively low. Therefore, the sender's willingness to pay for any implementable bundle is primarily determined by his value for $L$, which decreases in type. Although bundle $R^*$ offers more $q_R$ than $B$, it also offers more $q_L$. Thus, a lower interval of types $[\tstar,\tstaru)$ are willing to pay more for bundle $R^*$ than the higher interval $(\tstar,\thetah]$. By implication, $q_R(\cdot)$ is  non-monotone in the \unb{} environment ($q_L(\cdot)$ is monotone, as in balanced). Furthermore, the leftmost interval of types has highest willingness to pay to induce action $L$. To maintain a high price of test $L^*$ for this group, some left-leaning types must instead be allocated $R^*$. Despite this non-monotone pattern of distortions, the pattern of information rents is still monotone decreasing; see Figure \ref{Fig:utility}.

The think tank interpretation in this setting has some notable differences from the \bln{} environment. It is still true that the intermediary establishes three think tanks: two with overt agendas (who only offer $L^*$ and $R^*$) and a neutral think tank that offers an ``equalizing'' test. The first crucial difference is that the equalizing test is no longer $(\half,\half)$. While it leads to identical probability of actions $L$ and $R$, it also induces a positive probability of action $\s$. In other words, the neutral think tank sometimes fails to persuade the receiver to switch from the default action. This seemingly minor change has important implications, not only for the prices each think tank charges, but also for the matching between senders and think tanks. While in the \bln{} environment the matching was assortative, in the \unb{} environment a reversal occurs: middle types, including some who lean left, select think tank $R$, while extreme right types employ the (ex post) neutral think tank. Thus, senders who favor $R$ most intensely are less likely to induce it than more moderate types, including some who lean left.

\subsection{Changes in the Prior Belief}
We examine how changes in the prior belief affect the optimal screening menu and the sender's indirect utility. Consider a decrease in the prior belief $\pri$, small enough that the environment does not switch from balanced to unbalanced.\footnote{A drop in $\pri$ could not possibly switch the environment from unbalanced to balanced.} With such a decrease, state $0$ is more likely. Consequently, it is more difficult to persuade the receiver to select action $R$ and less difficult to persuade the receiver to select $L$. Naturally, the implementable set $Q$ shifts southeast: extreme points $L^*$ and $R^*$, shift down and to the right ($q_L$ increases, while $q_R$ decreases), while $R_0^*$ shifts down and $L_0^*$ shifts right. This shift has an interesting property: the slopes of the faces of $Q$ are not affected by the change (consult \eqref{slopes}). It follows that, in both the \bln{} and \unb{} environments, the cutoff types $\tstar,\tstaru,\tstarb$ are not affected by the change in the prior belief (as is evident from their characterization in \eqref{eq:cut1} and \eqref{eq:cut3}). Indeed, these cutoffs types are determined by a comparison between the slope of the indifference curves of the intermediary's objective and the slope of the faces of the implementable set.

The southeastern translation of $L^*$ and $R^*$ generates a clockwise rotation of the sender's indirect utility $U(\cdot)$. Recall that \eqref{eq:INT} implies that the slope of sender's indirect utility is $U'(\cdot)=q_R(\cdot)-q_L(\cdot)$. It follows that for types who receive $L^*$, the southeastern translation of the extreme points makes $U'(\cdot)$ even more negative, producing a steeper indirect utility. For types that receive $R^*$, the slope of indirect utility also decreases. In the balanced case, this slope is positive, so the reduction generates a flatter (less positively sloped) indirect utility. In the unbalanced case, this slope is negative, so the reduction generates a steeper (more negatively sloped) indirect utility. Combined with the observations that the thresholds $\tstar,\tstarb,\tstaru$ do not change with $\mu_0$, and $U(\cdot)$ must be continuous, these arguments suffice to prove the following proposition, illustrated in Figure \ref{Fig:utility_comp_statics}.

\begin{proposition}[Changes in Prior Belief] Consider a reduction in $\pri$, the prior probability of $\omega=1$, which is small enough that the environment does not switch from \bln{} to \unb{}. 

\begin{enumerate}
\item[(i)] Thresholds $\tstar,\tstarb,\tstaru$ are not affected by the reduction.
\item[(ii)] In a \bln{} environment, such a reduction leaves  types who receive $L^*$ ($\theta<\tstar$) better off, types who receive $R^*$ ($\theta>\tstarb$) worse off, and it does not affect the utility of types who receive $(\half,\half)$.
\item[(iii)] In an \unb{} environment, such a reduction leaves types who receive $L^*$ or $R^*$ ($\theta<\tstaru$) better off, and it does not affect the utility of types who receive $B$. If $\tstaru>\half$, then an interval of right-leaning types benefits when the prior probability of $\omega=1$ decreases.
\end{enumerate}  
\end{proposition}

\begin{figure}
\begin{minipage}{0.5\textwidth}
\begin{tikzpicture}[scale=0.5]


\draw[line width=0.5pt] (0,0) -- (0,5) ;
\draw[line width=0.5pt] (0,0) -- (15,0); 
\fill (15,0) node[right] {\footnotesize{$\theta$}};

\draw[line width=0.5pt,blue] (15,2.5) -- (10,0) -- (2.5,0) -- (0,.833); 
\draw[line width=0.5pt,purple] (10,0) -- (2.5,0); 


\draw[line width=0.5pt,red] (15,1) -- (10,0);
\draw[line width=0.5pt,red]  (2.5,0) -- (0,1.5);




\filldraw[color=blue, fill=white] (10,0) circle (3pt);
\node[below] at (10,0) {$1$};


\filldraw[color=blue, fill=white] (2.5,0) circle (3pt);

\node[below] at (2.5,0) {$\frac{1}{4}$};

\fill (1.25,0) node[below] {\footnotesize{$L^*$}};
\fill (6.25,0) node[below] {\footnotesize{$(\frac{1}{2},\frac{1}{2})$}};
\fill (12.5,0) node[below] {\footnotesize{$R^*$}};

\end{tikzpicture}
\end{minipage}
 \begin{minipage}{0.5\textwidth}
\begin{tikzpicture}[scale=0.5]

 \draw[line width=0.5pt] (0,0) -- (0,5); 
\draw[line width=0.5pt] (0,0) -- (15,0); 
\fill (15,0) node[right] {\footnotesize{$\theta$}};

\draw[line width=0.5pt,blue] (15,0) -- (12.5,0) -- (2.5,1) -- (0,3.167); 


\draw[line width=0.5pt,red] (12.5,0) -- (2.5,1.75) -- (0,4.083); 
\filldraw[color=red, fill=white] (2.5,1.75) circle (3pt);


\filldraw[color=blue, fill=white] (12.5,0) circle (3pt);
\node[below] at (12.5,0) {$\frac{5}{4}$};


\draw[line width=0.5pt, dotted] (2.5,0) -- (2.5,1.75); 
\filldraw[color=blue, fill=white] (2.5,1) circle (3pt);

\node[below] at (2.5,0) {$\frac{1}{4}$};

\fill (1.25,0) node[below] {\footnotesize{$L^*$}};
 \fill (7.5,0) node[below] {\footnotesize{$R^*$}};
\fill (13.75,0) node[below] {\footnotesize{$B$}};

\end{tikzpicture}

\end{minipage}
\caption{The effect of a reduction in the prior $\mu_0$ on the sender's indirect utility $U(\cdot)$ under the optimal screening menu. Using the same parameters as Figure \ref{Fig:utility}, the left panel shows a reduction from $\pri=.5$ (blue) to $\pri=.4$ (red) in the balanced case, and the right panel shows a reduction from $\pri=.3$ (blue) to $\pri=.275$ (red) in the unbalanced case.\label{Fig:utility_comp_statics}}
\end{figure}

This proposition has a number of interesting aspects. Point (i) implies that the matching of senders to tests (or think tanks) depends on the distribution of sender types, but it is not affected by changes in the prior belief (``public opinion''), unless the ``balancedness'' of the environment is altered.   As described previously, this finding arises from a property of the (endogenous) implementable set $Q$, that the slopes of its faces do not depend on the prior belief. Point (ii) implies that in the \bln{} environment,  changes in the prior belief affect the sender in an intuitive way---as the prior belief shifts in favor of action $L$, sender types with the strongest vested interest in $L$ benefit, while types with the strongest vested in $R$ are harmed. 

In contrast, point (iii) reveals that the prior belief and  optimal screening menu interact counterintuitively when the environment is unbalanced---a shift in the prior belief in favor of action $L$ not only makes all sender types weakly better off, it may strictly benefit moderate right-leaning types. To understand this point, note that the southeastern shift in bundle $R^*$ that accompanies the reduction in the prior belief produces a clockwise rotation in the willingness to pay for $R^*$, around pivot point $(\half,\half)$. In particular, left-leaning sender  ($\theta<\half$) are more willing to pay for $R^*$ after the shift, since this bundle contains more $q_L$ and less $q_R$. Right-leaning types ($\theta>\half$) are less willing to pay for $R^*$ after the shift, but the drop in willingness to pay is larger for larger types, who have the strongest vested interest in action $R$. 
Thus, type $\tstaru$'s  willingness to pay for $R^*$ drops by more than that of any other type $\theta\in[\tstar,\tstaru)$ that buys test $R^*$. Because the price of $R^*$ in the optimal menu is simply type $\tstaru$'s willingness to pay, the drop in price is larger than the drop in willingness to pay for all such types, which increases their indirect utility. 

Of particular note, a shift in the prior that favors action $L$ may benefit some right-leaning types ($\theta\in[\half,\tstaru]$). This finding arises as a consequence of screening---if the sender controlled information provision, or if his type was observable by the intermediary, then all right-leaning types $(\theta>\half)$ would select $R^*$, and all such types would be harmed by the reduction in the prior belief.

\subsection{Intermediation and Receiver Welfare}\label{intermediation} We examine the intermediary's impact on information transmission and the receiver's welfare. In particular, we compare the receiver's welfare in two settings: (1) when information is generated by the intermediary's optimal screening menu (i.e.,  ``screening''), and (2) when each sender type transmits the result of his payoff-maximizing test (``unmediated communication''). This version of unmediated communication can arise from a variety of changes to the environment. For example, if adjustments to disclosure rules allow the intermediary to observe the sender's type, then the intermediary would offer the first-best menu, which maximizes (and extracts) each sender type's surplus. Alternatively, a merger of sender and intermediary would also give sender greater authority over information production. Competition between senders might also allow sender the freedom to select the test that he prefers. Indeed, in the case of perfect competition, sender can transmit his most-preferred test at zero marginal cost.

To ease the exposition and highlight the most interesting case,  we focus on $\mul<\half<\muh$ and the balanced environment ($\muh<2\pri<1+\mul$); in anticipation of results, note that an interval of prior beliefs around $\half$ always satisfies both of these conditions. In contrast to the first best, in the optimal screening menu the types in $(\tstar,\half)$ select \infl{} bundle $(\half,\half)$ instead of $L^*$, and types in $(\half,\tstarb)$ select $(\half,\half)$ instead of $R^*$. To evaluate the normative impact on the receiver, we must therefore compare the receiver's welfare when \infl{} bundle ($\half,\half$) replaces $R^*$ and $L^*$. The analysis is complicated by the fact that many tests implement \infl{} bundle $(\half,\half)$; while sender and intermediary are indifferent among them, the receiver is not. To address this issue, we characterize the receiver's preferred test among these, and focus on the (Pareto Optimal) equilibrium in which the intermediary induces $(\half,\half)$ with the test that maximizes the receiver's payoff, which we refer to as the \textit{receiver-optimal equalizing test}.

\begin{lemma}[Receiver-Optimal Equalizing Test]\label{lem:rec_opt} Suppose the environment is \bln{}. Among all obedient decision rules that induce \infl{} bundle $(\half,\half)$, the one that maximizes the receiver's payoff
\begin{enumerate}
    \item[(i)] generates posterior belief $\mu_R=2\pri$ when recommendation $R$ is issued and posterior belief $\mu_L=0$ when $L$ is issued, if  $\pri<\half$. 
    \item[(ii)] generates posterior belief $\mu_R=1$ when recommendation $R$ is issued and posterior belief $\mu_L=2\pri-1$ when $L$ is issued, if  $\pri>\half$. 
    \item[(iii)] is fully informative if and only if $\pri=\half$.
\end{enumerate}

\end{lemma}

Intuitively, among all tests that induce \infl{} bundle $(\half,\half)$, we seek the one in which the posterior beliefs are most spread out. Furthermore, because the equalizing \infl{} bundle is induced, the posterior beliefs must average to the prior when weighted equally. 
Whether $\mu_0$ is below or above $\half$ determines whether the  constraint $\mul\geq 0$ or the constraint $\muh\leq 1$ binds first. 
In the symmetric case, $\pri=\half$, these constraints bind simultaneously, resulting in a fully revealing test.

The receiver-optimal equalizing test is continuous in the prior belief, as is the receiver's payoff. By implication, for an interval of prior beliefs near $\half$, the payoff of the receiver-optimal equalizing test is close to the full information payoff, which is strictly higher than the payoff from both $R^*$ and $L^*$. It follows immediately by continuity that for prior beliefs near $\frac{1}{2}$, a monopolist intermediary is better than a perfectly competitive market for the receiver regardless of the distribution of sender types.\footnote{For $\mu_0$ farther away from $\frac{1}{2}$, whether the receiver still prefers the screening menu depends on the distribution of types. For such priors, the receiver-optimal equalizing test is worse for the receiver than one of the first-best tests and better than the other. The distribution of types governs the tradeoff between these gains and losses.} The following result requires no further proof.

\begin{proposition}[Mediated vs. Unmediated Communication]\label{prop:screening_vs_transparency} Suppose the environment is \bln{} and $\mul<\half<\muh$. For each $\{\mul,\muh\}$, there exists an interval $(\underline{\mu}^*,\overline{\mu}^*)$ surrounding $\half$ such that for all $\pri\in(\underline{\mu}^*,\overline{\mu}^*)$,  the receiver's payoff is higher when the intermediary screens the sender than with unmediated communication, regardless of the distribution of sender types.
\end{proposition}

This proposition shows that unmediated communication,  whereby the sender's payoff-maximizing test is always transmitted, can lower the receiver's welfare. Thus, changes to disclosure rules that increase transparency, changes in market structure that increase sender's bargaining power, or mergers that increase sender's authority over test design may reduce the receiver's welfare.

\subsection{Recap of Qualitative Findings} 
Before considering extensions, we summarize our main qualitative findings. The properties of the optimal screening menu depend critically on whether the environment is balanced or unbalanced, which in turn depends on the receiver's decision problem. We highlight some of the qualitative differences here.

\paragraph{Test Design.} In the \bln{} and \unb{} environments, the optimal screening menu contains the two efficient \infl{} bundles, $L^*$ and $R^*$, and a ``neutral '' bundle, which induces an equal probability of receiver selecting $L$ and $R$. In the \bln{} environment, the neutral bundle is $(\half,\half)$: receiver is always persuaded to act. In the \unb{} environment, the neutral bundle is $B$, and the probability that the receiver does not act (selects the status quo $S$) is positive.
    
    \paragraph{Matching.} In the \bln{} environment, types are matched assortatively to tests: only left-leaning types purchase $L^*$ and only right-leaning types purchasing $R^*$. In the \unb{} environment, matching is not assortative, with some left-leaning types selecting $R^*$, the test that maximizes the probability of action $R$.

    \paragraph{Rent Extraction.} In the \bln{} environment, the participation constraint binds on an interval around the neutral type, which includes both left- and right-leaning senders. Indirect utility increases as type becomes more extreme (further from $\half$). In the \unb{} environment the participation constraint binds on an interval of high types. Unless the distribution of types is (sufficiently) skewed or dispersed, this interval only contains right-leaning types. Furthermore, the sender's indirect utility is decreasing in type.

    \paragraph{Monotonicity of Action.} In the \bln{} environment, the probability of each action is monotone in type, $q_L(\cdot)$ is decreasing and $q_R(\cdot)$ is increasing. In the \unb{} environment, $q_R(\cdot)$ is non-monotone, first increasing, then decreasing. Extreme right-leaning types have a lower probability of persuading the receiver to select $R$ than moderate left-leaning types.
    
       \paragraph{Effect of the Prior.} In both environments, a change in the prior belief (that maintains the balanced/unbalanced classification) does not alter the matching of sender types to tests. In \bln{}, a shift down in the prior belief (favoring action $L$) increases the indirect utility of types that buy $L^*$ and reduces the indirect utility of types that buy $R^*$. In \unb{}, the shift increases the indirect utility of types that buy $L^*$ and $R^*$, including right-leaning types.
       
       \paragraph{Unmediated Communication.} Screening by the intermediary may improve the receiver's payoff. Each test in the optimal screening menu may be (weakly) more valuable to the receiver than the corresponding first-best test.

\section{Extensions}
In this section we study two variations of our model, adjusting the instruments that are available to the intermediary for inclusion in the screening menu. We provide a brief summary of each extension and the main findings here.

\paragraph{Coercion.} The first variation considers a stronger intermediary who can commit to providing information to the receiver even if the sender chooses not to participate in the mechanism, which changes each type's outside option. By doing so, the intermediary coerces types that dislike the outside option to pay more, but also provides some influence for free to other types, which reduces their willingness to pay. We provide a simple sufficient condition under which the optimal menu features coercion, provide a partial characterization, and show that the possibility of coercion increases the Blackwell-informativeness of  the tests in the optimal menu.

\paragraph{Selling Access.} The second variation makes the intermediary weaker: the sender controls information production, but the intermediary controls access to the receiver. In other words, if the sender pays for access, then he can design and carry out any test that he would like, but if he does not pay, then no information is conveyed. We identify the channels by which the intermediary's loss of control over information reduces her profit and the consequences for test informativeness.

\subsection{Coercion}\label{sec:coercion}
We now extend the model to allow the intermediary to commit to an influence bundle $(q_L(n),q_R(n))$ in case the sender chooses not to participate ($n$). Formally, the intermediary chooses a menu 
\begin{align*}
\mathcal{M}=\{(q_L(\theta),q_R(\theta),p(\theta))\}_{\theta\in\Theta}\cup \{(q_L(n),q_R(n))\}
\end{align*}
and asks the agent to make a report from $\Theta \cup \{n\}$, either reporting a type or opting out.\footnote{It would be counterproductive for the intermediary to offer multiple outside options in an attempt to tailor the threat to the agent's type, because the relevant outside option for each type is that type's favorite among all those offered. Given any (incentive compatible and individually rational) menu with multiple outside options, the intermediary could relax participation constraints by selecting any one of those outside options and making only that one available.} The intermediary maximizes the same objective \eqref{eq:objective}, subject to \eqref{eq:I}, \eqref{eq:IC}, and the new ``coercive" individual rationality constraints for all $\theta\in\Theta$:
\begin{align}
    u(q_L(\theta),q_R(\theta),\theta)-p(\theta) \geq  u(q_L(n),q_R(n),\theta).\label{eq:CIR}\tag{CIR-$\theta$}
\end{align}

Coercion through a bundle $(q_L(n),q_R(n))$ is only beneficial for the intermediary if it gives at least \textit{some} type negative utility, otherwise the intermediary could do just as well by setting $(q_L(n),q_R(n))=(0,0)$, which weakly relaxes \eqref{eq:CIR} and has no effect on the other constraints. A corollary is that coercion can only be (strictly) beneficial when either $\thetah>1$ or $\thetal<0$. That is, if $[\thetal,\thetah]\subseteq [0,1]$, then the menus described in Propositions \ref{prop:bln} and \ref{prop:unb} are optimal even within the larger set of menus that include coercion.\footnote{Recall our original assumption that $S$ is optimal under the prior. If $\pri>\muh$, for example, then the default action would be $R$, and even with $[\thetal,\thetah]\subseteq [0,1]$, it would be possible to coerce high types via information that induces $R$ with lower probability.}

To see the basic tradeoff, note that by introducing a coercive test, the intermediary may be able to extract more from extreme types who strongly dislike it. At the same time, by committing to reveal some information for free, the intermediary may undermine her ability to extract surplus from other types. For example, an increase in $q_L(n)$ reduces the value of the outside option for sufficiently extreme right-leaning types, but increases it for all other types. This change in the value of the outside option directly affects the price for types at which the participation constraint binds. At the same time, it may also shift the set of types at which the participation constraint binds, which indirectly affects the information rents for all other types through the incentive compatibility constraint. In general, identifying the optimal coercive bundle involves aggregating these direct and indirect effects across the type space, simultaneously optimizing both the ``inside'' and outside options.

Though complicated in general, when the intermediary observes the sender's type (first best), the characterization is relatively straightforward. Indeed, suppose that the intermediary can observe the type of any sender who decides to participate, but cannot observe the type of a sender who opts out. Thus, the intermediary must offer the same outside option $(q_R(n),q_L(n))$ to all types of sender.\footnote{If the intermediary could tailor the outside option to the sender's type, then the outside option would be set to minimize the sender's payoff type by type.} In this benchmark the intermediary can price every type's inside option so that his participation constraint  binds, so the intermediary only needs to account for the direct effect of changes to the outside option. In particular, the intermediary's objective function is 
\begin{align*}
    &\int_{\thetal}^{\thetah} f(\theta)[u(q_R(\theta),q_L(\theta),\theta)-u(q_R(n),q_L(n),\theta)]d\theta
\end{align*}
It is then apparent that the optimal ``inside option'' $(q_R(\theta),q_L(\theta))$ for each type is identical to the first best in the baseline model, while the outside option is designed to minimize the sender's ex ante expected utility. Because utility is linear in type, this is tantamount to minimizing the utility of the \textit{mean} type. Thus if the mean type inside the unit interval, then no coercion is used, but if it is sufficiently high, the coercive bundle is $L^*$ or $L_0^*$ (depending on parameter values); and similarly, if the mean type is sufficiently negative, the coercive bundle is $R^*$ or $R_0^*$.

When screening is a concern, the optimal outside option generally depends on fine-tuned properties of the type distribution.  Nevertheless, a simple condition allows us to construct a coercive bundle that improves the intermediary's profit. Furthermore, we provide a simple sufficient condition that ensures that the coercive bundle we construct is optimal. These conditions place minimal requirements on the distribution of types, as they depend only on the cutoff types $\tstaru,\tstarb$.

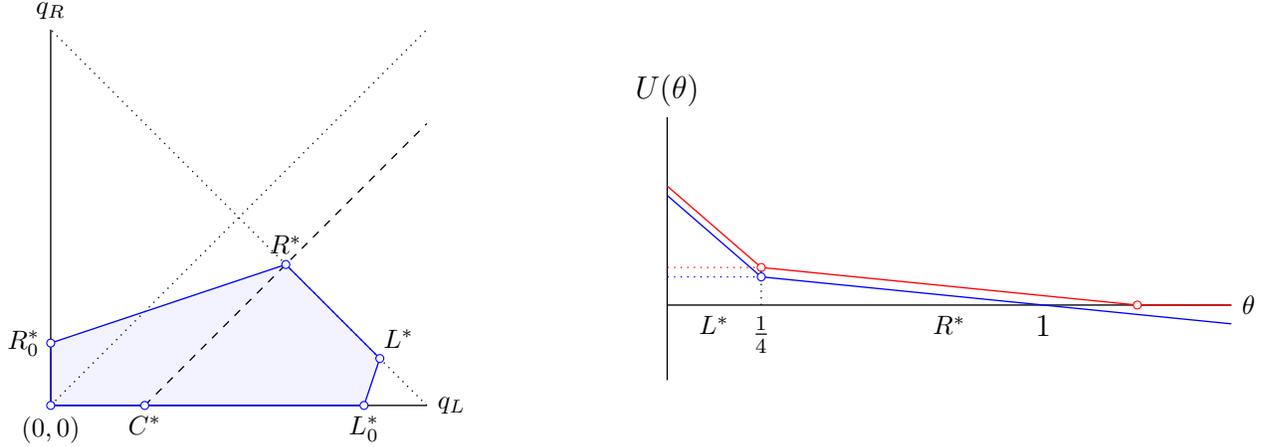
\begin{figure}
 \begin{minipage}{0.5\textwidth}
\begin{tikzpicture}[scale=0.5]
\draw[line width=0.5pt] (0,0) -- (0,10); 
\draw[line width=0.5pt] (0,0) -- (10,0); 
\fill (10,0) node[right] {\footnotesize{$q_L$}};
\fill (0,10) node[above] {\footnotesize{$q_R$}};


\draw[line width=0.5pt,dotted] (0,0) -- (10,10); 
\draw[line width=0.5pt,dotted] (10,0) -- (0,10); 

\draw[line width=0.5pt,dashed] (2.5,0) -- (10,7.5); 

\draw[line width=0.5pt,blue] (0,0) -- (0,1.66)--(6.25,3.75)--(8.75,1.25)--(8.33,0)--(0,0); 

 \filldraw[color=blue, fill=white] (0,0) circle (3pt);
\filldraw[color=blue, fill=white] (0,1.66) circle (3pt);
\filldraw[color=blue, fill=white] (6.25,3.75) circle (3pt);
\filldraw[color=blue, fill=white] (8.75,1.25) circle (3pt);
\filldraw[color=blue, fill=white] (8.33,0) circle (3pt);
\filldraw[color=blue, fill=white] (2.5,0) circle (3pt);

\fill (0,0) node[below] {\footnotesize{$(0,0)$}};
\fill (0,1.66)  node[left] {\footnotesize{$R^*_0$}};
\fill (6.25,3.75) node[above] {\footnotesize{$R^*$}};
\fill (2.5,0) node[below] {\footnotesize{$C^*$}};
\fill (9.25,1.25) node[above] {\footnotesize{$L^*$}};
\fill (8.33,0) node[below] {\footnotesize{$L_0^*$}};


\fill[blue,opacity=0.05] (0,0) -- (0,1.66)--(6.25,3.75)--(8.75,1.25)--(8.33,0)--(0,0);


\end{tikzpicture}
\end{minipage}
 \begin{minipage}{0.5\textwidth}
\begin{tikzpicture}[scale=0.5]


\draw[line width=0.5pt] (0,-2) -- (0,5) node[above]{$U(\theta)$}; 
\draw[line width=0.5pt] (0,0) -- (15,0); 
\fill (15,0) node[right] {\footnotesize{$\theta$}};

\draw[line width=0.5pt,blue] (15,-.5) --  (2.5,.75) -- (0,2.917); 
\draw[line width=0.5pt,blue,dotted] (2.5,.75) -- (0,0.75);
\draw[line width=0.5pt,red,dotted] (2.5,1) -- (0,1);

\draw[line width=0.5pt,red] (15,0) -- (12.5,0) -- (2.5,1) -- (0,3.167); 
\filldraw[color=red, fill=white] (12.5,0) circle (3pt);
\filldraw[color=red, fill=white] (2.5,1) circle (3pt);



\draw[line width=0.5pt, dotted] (2.5,0) -- (2.5,1); 
\filldraw[color=blue, fill=white] (2.5,.75) circle (3pt);

\node[below] at (2.5,0) {$\frac{1}{4}$};
\node[below] at (10,0) {$1$};

\fill (1.25,0) node[below] {\footnotesize{$L^*$}};
\fill (7.5,0) node[below] {\footnotesize{$R^*$}};


\end{tikzpicture}
\end{minipage}
\caption{\label{Fig:coercion}Proposition \ref{prop:coercion}. Left panel:  influence bundles offered under coercion $(L^*,R^*,C^*)$ within the implementable set. Right: sender's indirect utility under the optimal screening menu with coercion (blue) and without (red) for an unbalanced environment. Parameters from uniform example in Figure \ref{Fig:utility}, unbalanced environment. The optimal outside option is $C^*=(\frac{1}{10},0)$. Prices are $.45$ for $R^*$ and $.642$ for the $L^*$ test, both higher than their counterparts without coercion (Figure \ref{Fig:utility}). 
}
\end{figure}

To outline the construction, consider an \unb{} environment. Recall the optimal screening menu in Proposition \ref{prop:unb} and the sender's indirect utility plotted in the right panel of Figure \ref{Fig:utility}. Suppose, as in the figure, that $\tstaru\geq 1$, which implies $U(1)>0$. As a starting point, consider the possibility that the intermediary designs the outside option so that the sender's payoff from the outside option $u(q_R(n),q_L(n),\cdot)$ coincides with $U(\cdot)$ on interval $\theta\in[\tstar,\tstaru]$. To do so, the intermediary sets $q_R(n)=U(1)$ to ensure that the payoff of the outside option has the proper height. To ensure that the payoff of the outside option has the proper slope, the intermediary sets $q_R(n)-q_L(n)=2\frac{\pri}{\muh}-1$, which is the difference between $q_R$ and $q_L$ in bundle $R^*$.\footnote{Note that $u(q_R(n),q_L(n),\theta)=\theta q_R(n)+(1-\theta)q_L(n)$. Thus, both the the outside option and $U(\cdot)$ are linear on $[\tstar,\tstaru]$. For them to coincide on the interval, we must match their value at any point in the interval, which includes 1. Thus, we set $u(q_R(n),q_L(n),1)=q_R(n)=U(1)$. Furthermore, because types in this interval are allocated $R^*=(1-\pri/\muh,\pri/\muh)$, the slope of indirect utility is simply $U'(\theta)=q_R(\theta)-q_L(\theta)=2\pri/\muh-1$.} Crucially, this coercive bundle reduces the outside option for all $\theta\geq \tstaru$ without violating the participation constraint for any smaller type. Thus, in the (inside) menu, the intermediary can replace test $B$, intended for such types, with the more-profitable $R^*$ without violating participation (or incentive compatibility) constraints. The intermediary can make a further improvement by pushing the outside option's payoff down as far as possible without changing its slope. The intermediary can then increase prices equally for all types, shifting down the sender's indirect utility (without altering its shape) until the participation constraint binds. To shift down the outside option payoff in this way, the intermediary reduces $q_R(n)$ to 0 (reducing the height), while preserving the difference $q_R(n)-q_L(n)$ (maintaining the slope). The resulting coercive bundle is $C^*\equiv(1-2\frac{\pri}{\muh},0)$. This construction and the coercive bundle are illustrated in Figure \ref{Fig:coercion}.\footnote{If $\tstaru<1$ (i.e., $U(1)=0$), then no $(q_L(n),q_R(n))\in Q$ allows the intermediary to match the indirect utility on interval $[\tstar,\tstaru]$, since doing so would require $q_R(n)<0$.  In this case, the characterization depends on the exact tradeoff of payments between high and low types and might involve additional modifications to the inside options. }

\begin{proposition}[Coercion]\label{prop:coercion}
   Assume the environment is unbalanced.
   \begin{enumerate}
   \item[(i)] If $\theta^{**}_U>1$, then any optimal mechanism must feature coercion. 
   \item[(ii)] If $\theta^{**}_U>1$ and  $\theta^{**}_B\leq 1$, then it is optimal to coerce with outside option $C^*\equiv (1-2\frac{\pri}{\muh},0)$ and to offer two tests, $R^*$ to types above $\tstar$ and $L^*$ to types below. 
   \item[(iii)] If $\theta^{**}_U>1$ and  $\theta^{**}_B\leq 1$, then the test selected by each sender type in the optimal menu with coercion is (weakly) more Blackwell-informative than the test chosen by the same type without coercion. 
   
   \end{enumerate}
\end{proposition}  

Our characterization of the optimal coercive bundle reveals an interesting point: incentive compatibility constraints not only distort the inside options offered by the menu, they also distort the outside option. Indeed, the optimal outside option is not one of the extreme points, as it would be when a sender's type is observable. Finally, Proposition \ref{prop:coercion} shows that a stronger intermediary with the ability to coerce generates more information. To establish this formally, observe that the only change in the optimal allocation is that types who were assigned $B$ in the original screening menu are instead assigned $R^*$. It is straightforward to show that the test $B$ is a garbling of $R^*$.\footnote{$B$ is obtained from $R^*$ after replacing (i) a fraction $\frac{(\muh-\mul)(\muh-2\pri)}{(\muh-\pri)(\muh-2\mul)}$ of $L$ recommendations with $S$ and (ii) a fraction $\frac{\mul(\muh-2\pri)}{\pri(\muh-2\mul)}$ of $R$ recommendations with $S$.}

\subsection{Selling Access}

Our second extension disentangles two channels through which the (monopolist) intermediary extracts surplus, which are confounded in the main model. In particular, we weaken the intermediary by transferring control of the information production technology to the sender, but we still allow the intermediary to limit the receiver's \textit{access} to the information produced by the sender. Thus, the intermediary cannot design the test or experiment, but she can prevent the receiver from observing its realization unless the sender pays. To avoid additional complications, we assume that access must be purchased before the test outcome is realized. As an example, consider an online platform that allows firms, marketers, or political campaigns to display ads to targeted users in exchange for payment, without directly choosing or screening on the content of the ads. This benchmark allows us to distinguish the distortions generated by the intermediary's ability to control information content from those generated by her ability to limit access.

Suppose that the intermediary posts a price $p$ for access. A sender who does not pay cannot communicate with the receiver. Meanwhile, a sender who pays for access can design any test he likes, and the receiver observes the test and the recommendation before making its choice. Obviously, any sender who pays for access will implement his first-best bundle, $L^*$ for $\theta<\frac{1}{2}$ and $R^*$ for $\theta>\frac{1}{2}$, attaining the value depicted in Figure \ref{Fig:WTP}, while a sender who does not pay for access produces no information and attains value 0. Thus, the intermediary solves a monopoly pricing problem, where the good being sold is access.

Compared to a posted price, the optimal screening menu has two crucial differences. First, with a posted price, any sender types who choose not to buy access are completely excluded, whereas in the main model, the sender does not need to exclude any types and can instead offer the (receiver-optimal) equalizing test. Second, when selling access, the intermediary cannot make the price depend on the sender's chosen test. In the balanced environment the optimal screening menu (generally) sets different prices for tests $L^*$ and $R^*$ in order to exploit asymmetries in the distribution of sender types. Furthermore, in the unbalanced environment, the optimal screening menu sets such a low price for $R^*$ (relative to $L^*$) that some left-leaning senders purchase $R^*$. Obviously, neither of these pricing strategies can be implemented with a single price for access.

Weakening the intermediary also affects information quality and receiver welfare. In the unbalanced case, a weaker intermediary sets a price for access that excludes an interval of high types. In other words, an interval of high types transmits no information, rather than its optimal test from the screening menu, $L^*$, $R^*$, or $B$, all of which are informative and valuable to the receiver. Thus, in the unbalanced environment, a weaker intermediary generates less information about the state and reduces the receiver's welfare.  A similar argument applies to the balanced case, though it is more subtle. In the balanced case, the weaker intermediary excludes an interval of interior types. Thus every left-leaning (right-leaning) type provides either an uninformative test or $L^*$ ($R^*$). Under the optimal screening menu, every left-leaning (right-leaning) sender selects either $L^*$ ($R^*$) or the receiver-optimal equalizing test (see Lemma \ref{lem:rec_opt}). When the prior belief is near $\half$, the receiver-optimal equalizing test is more valuable to the receiver than $L^*$ and $R^*$. Thus, when the intermediary can only charge for access, every type's test is weakly less valuable to the receiver than in the optimal screening menu.

\section{Model Assumptions}\label{sec:discuss}
We discuss some of the assumptions of our model and how these could be relaxed.

\subsection{Costless Information Production}
As in the standard Bayesian persuasion framework, our intermediary incurs no costs for providing information. The advantage of this feature is that all distortions in the optimal menu are driven solely by screening considerations. However, it would also be natural to study a model in which information is costly depending on its precision, such as with a posterior-separable cost function.  To analyze such a model, for each influence bundle in the implementable set, one would characterize the test that induces it at lowest cost. The intermediary would incorporate this cost in her objective along with revenue. Depending on the specification of the cost functional, the indifference curves could be non-linear, and influence bundles along the faces of $Q$ could be offered as part of the optimal menu. However, if the costs of information production are small, the indifference curves are nearly linear, and we would expect the optimal menu to look similar to our characterization. The same approach could be used to study environments in which the intermediary has preferences over the receiver's action and the state, in addition to her revenue.

\subsection{Sender Payoffs}\label{dis:pay}

We present a natural microfoundation of the sender's preferences, embedding them into a standard model of vertical and horizontal differentiation. In particular, suppose that actions $\{L,\s, R\}$ represent policies with vertical qualities $\{v_L,0,v_R\}$, located in the standard single-dimensional policy space at $\{0,s,1\}$, with $s\in(0,1)$. A sender's type represents his location in the policy space, and the cost of mismatch is quadratic. With this specification,
\begin{align*}
    u(L,\theta)=v_L-\theta^2\qquad u(\s,\theta)=-\left(s-\theta\right)^2\qquad u(R,\theta)=v_R-(1-\theta)^2.
\end{align*}
Sender's utility net of the default action $\s$---which determines his willingness to pay for influence---is therefore
\begin{align*}
\tilde{u}(L,\theta)=v_L+s^2-2s\theta\qquad
   \tilde{u}(R,\theta)=v_R-1+s^2+2(1-s)\theta.
\end{align*}
Evidently, our exact specification is recovered setting $v_L=v_R=\frac{3}{4}$ and $s=\half$. Furthermore, for any specification of $(v_L,v_R,s)$, sender's payoffs are linear, monotone in opposite directions, but not necessarily constant sum.\footnote{Rather than a preference parameter, we could also interpret $\theta$ as sender's private belief about the payoff consequences of the receiver's action, which is uncertain to the sender and inaccessible to the intermediary. This also generates a linear specification.} The analysis is largely similar in this case, except that the role of the 45\textdegree{} line is played by a line with a different slope.

\subsection{Type Bounds}\label{sec:bounds}
Assumption \ref{a:bounds} is sufficient to ensure that there were no types so extreme that their first-best bundle is $L^*_0$ or $R^*_0$. However, the logic of Propositions \ref{prop:bln} and \ref{prop:unb} readily extend when this assumption is relaxed. To illustrate, suppose that there exist extremely high types violating Assumption \ref{a:bounds}. If $\kappa_R<1$ and there exists $\theta^\dag$ such that $\phi^+(\theta^\dag)=\frac{1}{1-\kappa_R}$ then the optimal menu would be expanded to offer bundle $R^*_0$ for types above $\theta^\dag$, while the rest of the menu would be characterized as in Proposition \ref{prop:bln} or \ref{prop:unb}. Likewise, if $\kappa_L>1$ and there exists $\theta^\ddag$ such that $\phi^-(\theta^\ddag)=-\frac{1}{\kappa_L-1}$, then the menu would be expanded to offer $L^*_0$ to all types below $\theta^\ddag$. The rest of the menu is unchanged (i.e., for types consistent with (A\ref{a:bounds})) because their allocations are determined by pointwise maximization in the relaxed problem, subject to \eqref{eq:IR-theta_up} and \eqref{eq:IR-theta_down}.

\section{Conclusion}

We have studied the design problem of an intermediary who has exclusive control over the information available to a decision maker. This privileged position allows the intermediary to extract surplus from interested parties who seek to influence the decision maker's choice. 

We formulate the intermediary's design problem as  a form of monopolistic screening, reflecting three distinctive features of our setting. First, the decision maker has a multi-faceted decision with heterogeneous effects. In our monopolistic screening problem, this manifests as a multi-dimensional allocation, horizontally differentiated buyer preferences, and countervailing incentives. Second, the intermediary can only exert influence by providing information. This manifests as technological (rather than incentive) constraints on the set of alternatives that the intermediary can offer. These features interact to shape the intermediary's optimal menu, producing an intriguing set of qualitative findings.

\section{Appendix}
This appendix contains the proofs of all results except Proposition \ref{prop:coercion} and Lemma \ref{lem:rec_opt}, which are proved in the Supplementary Appendix.
\subsection{Proof of Lemma \ref{lem:feasible_q_pairs}}
As described in \citet{BM2016}, any test can be represented as an obedient decision rule. Let $\pi_i(\omega)$, $i\in A$, be the probability that recommendation $i$ is issued in state $\omega$, and let $q_i=\pri\pi_i(1)+(1-\pri)\pi_i(0)$ be the ex ante probability.  Let $\mu_i\equiv\Pr(\omega=1|i)$ be the receiver's posterior belief that the state is 1 given recommendation $i\in A$. If $q_i>0$, then this belief is derived from Bayes' rule; otherwise, it can take any value. In particular,
\begin{align*}
    \mu_i=\frac{\pri\pi_i(1)}{q_i}\quad \text{if}\quad q_i>0\quad \text{and}\quad\mu_i\in[0,1]\quad \text{otherwise}.
\end{align*}
Because $q_L+q_R+q_\s=1$, at least one $q_i>0$. Therefore, for any decision rule,
\begin{align}\label{LIE}
    q_L\mu_L+q_R\mu_R+q_\s\mu_\s=\pri.
\end{align}
Obedience requires that the receiver prefers to comply with any recommendation issued by the decision rule upon receiving it,
\begin{align*}
    q_R>0\Rightarrow \mu_R\geq \muh\quad\quad
    q_L>0\Rightarrow \mu_L\leq \mul\quad\quad
    q_S>0\Rightarrow \mul\leq\mu_\s\leq \muh.
\end{align*}
The first condition ensures compliance with $R$, the second with $L$, the third with $\s$.

\noindent\textit{Part 1: We show that an obedient decision rule satisfies (\ref{eq:feas1}-\ref{eq:feas3}).} Obviously, $q_i\geq 0$ for $i\in A$, and furthermore $q_L+q_R+q_\s=1$, which implies $q_L+q_R\leq 1$, proving (\ref{eq:feas1}). Combining the obedience conditions for $L$ and $\s$, feasibility constraint $\mu_R\leq 1$, and (\ref{LIE}), we have 
\begin{align*}
    q_L\mul+q_R+(1-q_L-q_R)\muh\geq \pri\implies 
    (\muh-\mul)q_L-(1-\muh)q_R\leq \muh-\pri,
\end{align*}
which is (\ref{eq:feas2}). Similarly, combining the obedience conditions for $R$ and $\s$, feasibility constraint $\mu_L\geq 0$ and (\ref{LIE}),
\begin{align*}
    q_R\muh+(1-q_L-q_R)\mul \leq\pri\implies
    q_R(\muh-\mul)-q_L\mul \leq\pri-\mul,
\end{align*}
which is (\ref{eq:feas3}).

\noindent\textit{Part 2. We show that any $(q_L,q_R)$ which satisfies (\ref{eq:feas1}-\ref{eq:feas3}) can be induced by some test.} Consider $(q_L,q_R)$ satisfying (\ref{eq:feas1}-\ref{eq:feas3}). Let 
\begin{align}
    G(t)\equiv  q_L(t\mul) + (1-q_L-q_R)((1-t)\mul+t\muh)+q_R((1-t)\muh+t).
\end{align}
We have
    $G(0)=(1-q_L-q_R)\mul+q_R\muh\leq\pri$ by \eqref{eq:feas3}, and similarly, 
$G(1)=q_L\mul+(1-q_L-q_R)\muh+q_R\geq\pri$
by \eqref{eq:feas2}. 
By the intermediate value theorem, there exists $t^*\in [0,1]$ such that $G(t)=\pri$. By construction, the distribution of posteriors
\begin{align*}
    \mu_L=t^*\mul, \qquad \mu_\s=(1-t^*)\mul+t^*\muh,\qquad
    \mu_R=(1-t^*)\muh+t^*,
\end{align*}
with $\Pr(\mu_L)=q_L$, $\Pr(\mu_\s)=1-q_R-q_L$, and $\Pr(\mu_R)=q_R$, is Bayes-Plausible. By \citet{KG2011}, there exists a statistical experiment that induces it. Furthermore, by construction $\mu_L\leq \mul$, $\mu_\s\in[\mul,\muh]$, and $\mu_R\geq\muh$. By implication, an optimal strategy for receiver is to select action $i\in A$ if and only if belief $\mu_i$ is realized. Thus, the probability that receiver chooses action $i\in\{L,R,\s\}$ is $q_i$.

\subsection{Proof of Observation \ref{obs:FB}}

\begin{figure}
    \centering
    \begin{tikzpicture}[scale=0.7]

\draw[line width=0.5pt] (0,0) -- (0,10); 
\draw[line width=0.5pt] (0,0) -- (10,0); 
\fill (10,0) node[right] {\footnotesize{$q_L$}};
\fill (0,10) node[above] {\footnotesize{$q_R$}};

\draw[line width=0.5pt,dotted] (0,0) -- (10,10); 
\draw[line width=0.5pt,dotted] (10,0) -- (0,10); 

\draw[line width=0.5pt,blue] (0,0) -- (0,4.16)--(4.375,5.625)--(6.875,3.125)--(5.833,0)--(0,0); 

\draw[line width=0.5pt,red] (3.375,5.925)--(4.375,5.625)--(5.375,5.325); 
\draw[line width=0.5pt,red] (6.875-0.3,3.125+1)--(6.875,3.125)--(6.875+0.3,3.125-1);
\draw[line width=0.5pt,green] (3.375,5.525)--(4.375,5.625)--(5.375,5.725); 
\draw[line width=0.5pt,green] (0.5,5.16)--(0,4.16)--(-0.5,3.16);

 \filldraw[color=blue, fill=white] (0,0) circle (3pt);
\filldraw[color=blue, fill=white] (0,4.16) circle (3pt);
\filldraw[color=blue, fill=white] (4.375,5.625) circle (3pt);
\filldraw[color=blue, fill=white] (6.875,3.125) circle (3pt);
\filldraw[color=blue, fill=white] (5.833,0) circle (3pt);

\fill (0,0) node[above right,xshift=.7em] {\footnotesize{$45^\circ$}};

\fill (0,0) node[below] {\footnotesize{$(0,0)$}};
\fill (0,4.16) node[left] {\footnotesize{$R^*_0$}};
\fill (4.38,5.625) node[above] {\footnotesize{$R^*$}};
\fill (7.2,3.13) node[above] {\footnotesize{$L^*$}};
\fill (5.833,0) node[below] {\footnotesize{$L_0^*$}};

\fill (2,5.7) node[below] {\footnotesize{$(\ref{eq:feas3})$}};
\fill (6,4.5) node[above] {\footnotesize{$(\ref{eq:feas1})$}};
\fill (7,1) node[above] {\footnotesize{$(\ref{eq:feas2})$}};

\fill[blue,opacity=0.05] (0,0) -- (0,4.16)--(4.375,5.625)--(6.875,3.125)--(5.833,0)--(0,0);

\node[rotate=20] at (2,4.5) {\footnotesize{slope $\kappa_R$}};
\node[rotate=315] at (5.5,4) {\footnotesize{slope $-1$}};
\node[rotate=70] at (6,1.5) {\footnotesize{slope $\kappa_L$}};


\end{tikzpicture}

    \caption{Proof of Observation \ref{obs:FB}}
    \label{fig:PropFB}
\end{figure}

In the first best, the payment is set so that IR binds for every type. Thus the optimal menu $(q_L(\theta),q_R(\theta))$ maximizes $(1-\theta)q_L(\theta)+\theta q_R(\theta)$ subject to $(q_L(\theta),q_R(\theta))\in Q$. 

The most direct proof is graphical. In Figure \ref{fig:PropFB} the feasible set $Q$ is shaded blue. The intermediary's objective function has slope $\Delta\equiv-(1-\theta)/\theta$. Furthermore, at least one of $\{\theta,1-\theta\}$ is positive, thus the objective function increases along at least one axis. 

\noindent\textit{Case 1: $\theta\in[0,1]$}. If $\theta\in[0,1]$, then $\theta\geq 0$, $1-\theta\geq 0$, and one at least one inequality is strict. Thus, the objective function increases to the northeast. The optimum is at $R^*$ if $\Delta>-1$ (the indifference curve is flatter than face (\ref{eq:feas1})) and it is $L^*$ if $\Delta<-1$ (indifference curve is steeper). If $\Delta=-1$, then $R^*$, $L^*$ or any convex combination of them is optimal (see red lines in Figure \ref{fig:PropFB}). Therefore,
\begin{itemize}
\item   If $\theta\in[0,\half]$, then the first best \infl{} bundle is $L^*$.
\item If $\theta\in[\half,1]$, then the first best \infl{} bundle is $R^*$.
\end{itemize}

\noindent\textit{Case 2: $\theta\in[1,\thetah]$}. If $\theta\in[1,\thetah]$, then $\theta>0$ and $1-\theta\leq 0$. Thus, the objective function increases to the northwest and $\Delta>0$. The optimum is at $R^*$ if $\Delta<\kappa_R$ and is $R_0^*$ if $\Delta>\kappa_R$ (see green lines in Figure \ref{fig:PropFB}). Thus, $R^*$ is optimal whenever,
\begin{align*}
\Delta<\kappa_R\Leftrightarrow -\frac{1-\theta}{\theta}<\kappa_R    \Leftrightarrow\theta-1<\kappa_R\theta\Leftrightarrow \theta(1-\kappa_R)<1.
\end{align*}
If $\kappa_R\geq 1$, then the preceding holds trivially. If  $\kappa_R<1$, then it follows immediately from (A\ref{a:bounds}). Therefore, if $\theta\in[1,\thetah]$, then the first best \infl{} bundle is $R^*$.

\noindent\textit{Case 3:} $\theta\in[\thetal,0]$. This case is analogous to Case 2. The optimum is $L^*$ if $\Delta > \kappa_L$ and $L^*_0$ if $\Delta < \kappa_L$. The inequality $\Delta > \kappa_L$ reduces to $1 > \theta(1-\kappa_L)$, which holds if $\kappa_L\leq 1$ or if $\kappa_L > 1$ and $\theta >-\frac{1}{\kappa_L-1}$, which in turn holds by (A\ref{a:bounds}).

\subsection{Proof of Lemma \ref{lem:theta_0}}
We prove the result first for balanced environments and then for unbalanced environments.

\textbf{Balanced environments.} Let $\mathcal{M}$ be a solution to the intermediary's problem, and let $U$ be the agent's utility of truthful reporting under $\mathcal{M}$. If $U(\half)=0$, we are done, so suppose that $U(\half)>0$. We construct a new menu $\mathcal{M}'$ that improves on $\mathcal{M}$ as follows. Set $(q^{\M'}_L(\half),q^{\M'}_R(\half),p^{\M'}(\half))=(\half,\half,\half)$. (Since the environment is balanced, the $(\half,\half)$ allocation is implementable.) For each $\theta<\half$ such that $q^{\M}_L(\theta)\leq q^{\M}_R(\theta)$, set $(q^{\M'}_L(\theta),q^{\M'}_R(\theta),p^{\M'}(\theta))=(\half,\half,\half)$. Moreover, for each $\theta>\half$ such that $q^{\M}_L(\theta)\geq q^{\M}_R(\theta)$, set $(q^{\M'}_L(\theta),q^{\M'}_R(\theta),p^{\M'}(\theta))=(\half,\half,\half)$. For all other $\theta$, set $(q^{\M'}_L(\theta),q^{\M'}_R(\theta))=(q^{\M}_L(\theta),q^{\M}_R(\theta))$, and define
    \begin{align*}
    p^{\M'}(\theta)=\begin{cases}
        u(q^{\M'}_L(\theta),q^{\M'}_R(\theta),\theta)-U^{\M'}(\frac{1}{2})-\int_{\frac{1}{2}}^\theta (q^{\M'}_R(t)-q^{\M'}_L(t))\,dt\qquad &\text{if $\theta > \half$}\\
             u(q^{\M'}_L(\theta),q^{\M'}_R(\theta),\theta)-U^{\M'}(\frac{1}{2})+\int_\theta^{\frac{1}{2}} (q^{\M'}_R(t)-q^{\M'}_L(t))\,dt\qquad &\text{if $\theta < \half$},
    \end{cases}
    \end{align*}
    where $U^{\M'}(\half)=0$. (This characterization of $p^{\M'}(\theta)$ holds for all $\theta$; in particular, it reproduces a price of $\half$ in the instances where we had directly set it to $\half$.) It is straightforward to verify that $\M'$ satisfies all the relevant constraints. 

Now IR must bind under $\M$ for some type(s). Consider two cases. (Note that either Case (i) or Case (ii) below must apply; if neither applies, then IR binds at some types both above and below $\half$, and convexity implies it binds for an interval of types that includes $\half$ itself.)

Case (i): IR only binds under $\M$ at types above $\half$. In this case, define $\theta_0>\half$ as the infimum of such types. The constraint \eqref{eq:IR-theta_down} implies that for all $\theta<\theta_0$, $q_R^{\M}(\theta)\leq q_L^{\M}(\theta)$, and the constraint \eqref{eq:IR-theta_up} implies that for all $\theta>\theta_0$, $q_R^{\M}(\theta)\geq q_L^{\M}(\theta)$. It follows that the allocations under $\M$ and $\M'$ coincide outside of $[\half,\theta_0]$, and on this interval, $q^{\M'}_L(\theta)=q^{\M'}_R(\theta)=\half$ and $q^{\M}_L(\theta)\leq q^{\M}_R(\theta)$. A straightforward implication is that $U^{\M'}(\theta_0)=0$. Note that prices under $\M$ satisfy 
    \begin{align*}
    p^{\M}(\theta)=\begin{cases}
        u(q^{\M}_L(\theta),q^{\M}_R(\theta),\theta)-U^{\M}(\frac{1}{2})-\int_{\frac{1}{2}}^\theta (q^{\M}_R(t)-q^{\M}_L(t))\,dt\qquad &\text{if $\theta > \half$}\\
             u(q^{\M}_L(\theta),q^{\M}_R(\theta),\theta)-U^{\M}(\frac{1}{2})+\int_\theta^{\frac{1}{2}} (q^{\M}_R(t)-q^{\M}_L(t))\,dt\qquad &\text{if $\theta < \half$},
    \end{cases}
    \end{align*}
    where $U^{\M}(\half)> 0$. 

We argue that $p^{\M'}(\theta) \geq p^{\M}(\theta)$ for all $\theta\in \Theta$. First, observe that $U^{\M'}(\half)< U^{\M}(\half)$. For types $\theta < \half$, the allocations under $\M$ and $\M'$ are identical, so the difference in prices is $p^{\M'}(\theta)-p^{\M}(\theta)=U^{\M}(\half)-U^{\M'}(\half)>0$.
 For types $\theta\in [\half,\theta_0]$, we have $p^{\M'}(\theta)=\half$, while $U^{\M}(\theta)\geq 0$ implies
 \begin{align*}
     p^{\M}(\theta)\leq u(q^{\M}_L(\theta),q^{\M}_R(\theta),\theta)&=\theta q^{\M}_R(\theta)+(1-\theta)q^{\M}_L(\theta)\\
     &\leq \frac{1}{2} q^{\M}_R(\theta)+\frac{1}{2}q^{\M}_L(\theta)\leq \frac{1}{2},
 \end{align*}
    where the second inequality uses that $\theta\geq \half$ and $q^{\M}_R(\theta)\leq $$q^{\M}_L(\theta)$. Intuitively, surplus is higher for the types $\theta\in [\half,\theta_0]$ under $\M'$ \textit{and} the intermediary extracts all of it.
    For types $\theta>\theta_0$, rewrite the prices with respect to reference type $\theta_0$:
    \begin{align*}
        p^{\M}(\theta)&=u(q^{\M}_L(\theta),q^{\M}_R(\theta),\theta)-U^{\M}(\theta_0)-\int_{\theta_0}^\theta (q^{\M}_R(t)-q^{\M}_L(t))\,dt\\
        p^{\M'}(\theta)&=u(q^{\M'}_L(\theta),q^{\M'}_R(\theta),\theta)-U^{\M'}(\theta_0)-\int_{\theta_0}^\theta (q^{\M'}_R(t)-q^{\M'}_L(t))\,dt.
    \end{align*}
    By construction, the allocations coincide for $\theta>\theta_0$, so the difference in prices is
    \begin{align*}
        p^{\M'}(\theta)-p^{\M}(\theta)=U^{\M}(\theta_0)-U^{\M'}(\theta_0)=0-0=0.
    \end{align*}

Case (ii): IR only binds under $\M$ at types below $\half$. The arguments are analogous to those in case (i). Specifically, define $\theta_0<\half$ as the supremum of types at which IR binds under $\M$. By construction, allocations under $\M$ and $\M'$ are identical except for $\theta \in [\theta_0,\half]$. We again prove that $p^{\M'}(\theta)\geq p^{\M}(\theta)$ for $\theta<\theta_0$, $\theta\in [\theta_0,\half]$, and $\theta > \half$. Let us first address $\theta\geq \half$. Note that, as in Case (i), $U^{\M'}(\half)=0< U^{\M}(\half)$. Since allocations are identical under $\M$ and $\M'$ for types $\theta>\half$, the difference in prices is $p^{\M'}(\theta)-p^{\M}(\theta)=U^{\M}(\half)-U^{\M'}(\half)> 0$. Next, for $\theta \in [\theta_0,\half]$, we have $p^{\M'}(\theta)=\half$ while $p^{\M}(\theta)\leq \half$ by the same argument used in Case (i) for $\theta\in [\half,\theta_0]$. Finally, for $\theta <\theta_0$, one can easily establish that prices are the same by an argument analogous to the one for Case (i) when $\theta > \theta_0$: the prices calculated using reference type $\theta_0$ and the allocations for these types are identical under the two menus, and moreover, $U^{\M'}(\theta_0)=U^{\M}(\theta_0)=0$.

\textbf{Unbalanced environments.} 
 Suppose that under $\M$, the highest type at which IR binds is $\theta_0< \overline\theta$. Define a modified menu $\M'$ as follows. For all $\theta\geq \theta_0$, set $(q^{\M'}_L(\theta),q^{\M'}_R(\theta))=B$ and set $p^{\M'}(\theta)=\frac{\pri-\underline\mu}{\overline\mu-2\underline\mu}.$ For all $\theta<\theta_0$, set $(q^{\M'}_L(\theta),q^{\M'}_R(\theta),p^{\M'}(\theta))=(q^{\M}_L(\theta),q^{\M}_R(\theta),p^{\M}(\theta))$. Obviously, $\M'$ satisfies implementability. Since $\M$ satisfies monotonicity, $\M'$ satisfies monotonicity up to $\theta_0$. And since $q^{\M'}_R(\theta)-q^{\M'}_L(\theta)=0$ for all $\theta>\theta_0$ and \eqref{eq:IR-theta_down} (applied to $\M$) implies that for all $\theta<\theta_0$ we have $q^{\M'}_R(\theta)-q^{\M'}_L(\theta)=q^{\M}_R(\theta)-q^{\M}_L(\theta) \leq 0$, $\M'$ satisfies monotonicity. We claim that $\M'$ satisfies 
        \begin{align}
    p^{\M'}(\theta)=\begin{cases}
        u(q^{\M'}_L(\theta),q^{\M'}_R(\theta),\theta)-U^{\M'}(\theta_0)-\int_{\theta_0}^\theta (q^{\M'}_R(t)-q^{\M'}_L(t))\,dt\qquad &\text{if $\theta > \theta_0$}\\
             u(q^{\M'}_L(\theta),q^{\M'}_R(\theta),\theta)-U^{\M'}(\theta_0)+\int_\theta^{\theta_0} (q^{\M'}_R(t)-q^{\M'}_L(t))\,dt\qquad &\text{if $\theta < \theta_0$},
    \end{cases}
    \end{align}
   with $U^{\M'}(\theta_0)=0=U^{\M}(\theta_0)$. For $\theta<\theta_0$, this property is inherited from $\M$, and for $\theta>\theta_0$, the right hand side reduces to $\frac{\pri-\underline\mu}{\overline\mu-2\underline\mu}-0-\int_{\theta_0}^\theta 0\,dt=p^{\M'}(\theta)$. Thus, $\M'$ satisfies implementability, (IR), and (IC). 
    
We now show that $\mathcal{M}'$ improves on $\mathcal{M}$ by showing that for all $\theta\in \Theta$, $p^{\M'}(\theta)\geq p^{\M}(\theta)$, with strict inequality for $\theta> \theta_0$. Equality holds for $\theta\leq\theta_0$, so consider $\theta > \theta_0$. Since     $p^{\M'}(\theta)=u(q^{\M'}_L(\theta),q^{\M'}_R(\theta),\theta)$ and $p^{\M}(\theta)\leq u(q^{\M}_L(\theta),q^{\M}_R(\theta),\theta)$ by \eqref{eq:IR}, it is enough to show that $u(q^{\M'}_L(\theta),q^{\M'}_R(\theta),\theta)> u(q^{\M}_L(\theta),q^{\M}_R(\theta),\theta)$. By \eqref{eq:IR-theta_up}, $q^{\M}_R(\theta)\geq q^{\M}_L(\theta)$. Furthermore, since $\theta_0$ is by definition the highest type at which IR binds, \eqref{eq:INT} and \eqref{eq:MON} imply $q^{\M}_R(\theta)> q^{\M}_L(\theta)$ for all $\theta>\theta_0$. Thus, $(q^{\M}_L(\theta),q^{\M}_R(\theta))$ lies in the portion of the feasible set strictly above the 45\textdegree{} line. The closure of this set has 3 extreme points; one is $(q^{\M'}_L(\theta),q^{\M'}_R(\theta))=B$ and the others are $(0,0)$ and $R^*_0$. The allocation $B$ is the unique optimum by Assumption \ref{a:bounds}, and since $(q^{\M}_L(\theta),q^{\M}_R(\theta))$ lies strictly above the 45\textdegree{} line, $u(q^{\M'}_L(\theta),q^{\M'}_R(\theta),\theta))>u(q^{\M}_L(\theta),q^{\M}_R(\theta),\theta)$, concluding the proof.

\subsection{Proof of Proposition \ref{prop:bln}}
Let the \textit{lower optimization} problem be 
\begin{align*}
\max_{(q_L(\cdot),q_R(\cdot))}\int_{\thetal}^{\half}f(\theta)u(q_L(\theta),q_R(\theta),\phi^-(\theta))d\theta
\qquad\text{subject to \eqref{eq:I}, \eqref{eq:IR-theta_down}}. 
\end{align*}
Similarly, let the \textit{upper optimization} problem be
\begin{align*}
\max_{(q_L(\cdot),q_R(\cdot))}\int_{\half}^{\thetah}f(\theta)u(q_L(\theta),q_R(\theta),\phi^+(\theta))d\theta
\qquad\text{subject to \eqref{eq:I}, \eqref{eq:IR-theta_up}}. 
\end{align*}
Let $(q_L^-(\theta),q_R^-(\theta))$ and $(q_L^+(\theta),q_R^+(\theta))$ be the pointwise solutions of the lower and upper optimizations, respectively, both to be determined.

First, we solve the lower and upper optimization. We then show that the optimal menu for the relaxed program satisfies (MON), and therefore it solves the intermediary's problem.

\noindent\textit{Step 1. We solve the lower optimization.} The lower optimization can be done pointwise, $\max_{(q_L(\theta),q_R(\theta))}u(q_L(\theta),q_R(\theta),\phi^-(\theta))$, subject to $(q_L(\theta),q_R(\theta))\in Q^-\equiv Q\cap \{q_L(\theta)\geq q_R(\theta)\}$.

\begin{figure}
    \centering
     \begin{minipage}{0.45\textwidth}
    \begin{tikzpicture}[scale=0.65]

\draw[line width=0.5pt] (0,0) -- (0,10); 
\draw[line width=0.5pt] (0,0) -- (10,0); 
\fill (10,0) node[right] {\footnotesize{$q_L$}};
\fill (0,10) node[above] {\footnotesize{$q_R$}};

\draw[line width=0.5pt,dotted] (0,0) -- (10,10); 
\draw[line width=0.5pt,dotted] (10,0) -- (0,10); 

\draw[line width=0.5pt,gray] (0,0) -- (0,4.16)--(4.375,5.625)--(5,5)--(0,0); 
\draw[line width=0.5pt,blue] (0,0) -- (5,5)--(6.875,3.125)--(5.833,0)--(0,0);

\draw[line width=0.5pt,red] (5-1,5+0.3)--(5,5)--(5+1,5-0.3); 
\draw[line width=0.5pt,red] (6.875-0.3,3.125+1)--(6.875,3.125)--(6.875+0.3,3.125-1);
\draw[line width=0.5pt,green] (5-1,5-0.5)--(5,5)--(5+1,5+.5);

 \filldraw[color=gray, fill=white] (0,0) circle (3pt);
\filldraw[color=gray, fill=white] (0,4.16) circle (3pt);
\filldraw[color=gray, fill=white] (4.375,5.625) circle (3pt);
\filldraw[color=blue, fill=white] (6.875,3.125) circle (3pt);
\filldraw[color=blue, fill=white] (5.833,0) circle (3pt);
\filldraw[color=blue, fill=white] (5,5) circle (3pt);

\fill (0,0) node[above right,xshift=.7em] {\footnotesize{$45^\circ$}};

\fill (0,0) node[below] {\footnotesize{$(0,0)$}};
\fill (0,4.16) node[left] {\footnotesize{$R^*_0$}};
\fill (4.38,5.625) node[above] {\footnotesize{$R^*$}};
\fill (7.2,3.13) node[above] {\footnotesize{$L^*$}};
\fill (5.833,0) node[below] {\footnotesize{$L_0^*$}};
\fill (5.5,5) node[right] {\scriptsize{$(\half,\half)$}};


\fill[gray,opacity=0.05] (0,0) -- (0,4.16)--(4.375,5.625)--(5,5)--(0,0);

\fill[blue,opacity=0.05] (0,0) -- (5,5)--(6.875,3.125)--(5.833,0)--(0,0);

\node[rotate=20] at (2,4.5) {\footnotesize{slope $\kappa_R$}};
\node[rotate=315] at (5.5,4) {\footnotesize{slope $-1$}};
\node[rotate=70] at (6,1.5) {\footnotesize{slope $\kappa_L$}};
\node[rotate=45] at (2,1.6) {\footnotesize{slope $1$}};


\end{tikzpicture}
\end{minipage}
 \begin{minipage}{0.45\textwidth}
 \begin{tikzpicture}[scale=0.65]

\draw[line width=0.5pt] (0,0) -- (0,10); 
\draw[line width=0.5pt] (0,0) -- (10,0); 
\fill (10,0) node[right] {\footnotesize{$q_L$}};
\fill (0,10) node[above] {\footnotesize{$q_R$}};

\draw[line width=0.5pt,dotted] (0,0) -- (10,10); 
\draw[line width=0.5pt,dotted] (10,0) -- (0,10); 

\draw[line width=0.5pt,blue] (0,0) -- (0,4.16)--(4.375,5.625)--(5,5)--(0,0); 
\draw[line width=0.5pt,gray] (0,0) -- (5,5)--(6.875,3.125)--(5.833,0)--(0,0);

\draw[line width=0.5pt,red] (4.375-1,5.625+0.3)--(4.375,5.625)--(4.375+1,5.625-0.3); 
\draw[line width=0.5pt,red] (5+0.5,5-1)--(5,5)--(5-0.5,5+1);
\draw[line width=0.5pt,green] (5-0.5,5-1)--(5,5)--(5+0.5,5+1);
\draw[line width=0.5pt,green] (4.375-1,5.625-0.1)--(4.375,5.625)--(4.375+1,5.625+.1);

 \filldraw[color=blue, fill=white] (0,0) circle (3pt);
\filldraw[color=blue, fill=white] (0,4.16) circle (3pt);
\filldraw[color=blue, fill=white] (4.375,5.625) circle (3pt);
\filldraw[color=gray, fill=white] (6.875,3.125) circle (3pt);
\filldraw[color=gray, fill=white] (5.833,0) circle (3pt);
\filldraw[color=gray, fill=white] (5,5) circle (3pt);

\fill (0,0) node[above right,xshift=.7em] {\footnotesize{$45^\circ$}};

\fill (0,0) node[below] {\footnotesize{$(0,0)$}};
\fill (0,4.16) node[left] {\footnotesize{$R^*_0$}};
\fill (4.38,5.625) node[above] {\footnotesize{$R^*$}};
\fill (7.2,3.13) node[above] {\footnotesize{$L^*$}};
\fill (5.833,0) node[below] {\footnotesize{$L_0^*$}};
\fill (5.5,5) node[right] {\scriptsize{$(\half,\half)$}};


\fill[blue,opacity=0.05] (0,0) -- (0,4.16)--(4.375,5.625)--(5,5)--(0,0);

\fill[gray,opacity=0.05] (0,0) -- (5,5)--(6.875,3.125)--(5.833,0)--(0,0);

\node[rotate=20] at (2,4.5) {\footnotesize{slope $\kappa_R$}};
\node[rotate=315] at (5.5,4) {\footnotesize{slope $-1$}};
\node[rotate=70] at (6,1.5) {\footnotesize{slope $\kappa_L$}};
\node[rotate=45] at (2,1.6) {\footnotesize{slope $1$}};


\end{tikzpicture}
\end{minipage}
    \caption{Proof of Proposition \ref{prop:bln}: Lower optimization (left) and upper optimization (right).}
    \label{fig:PropBLN_proof}
\end{figure}
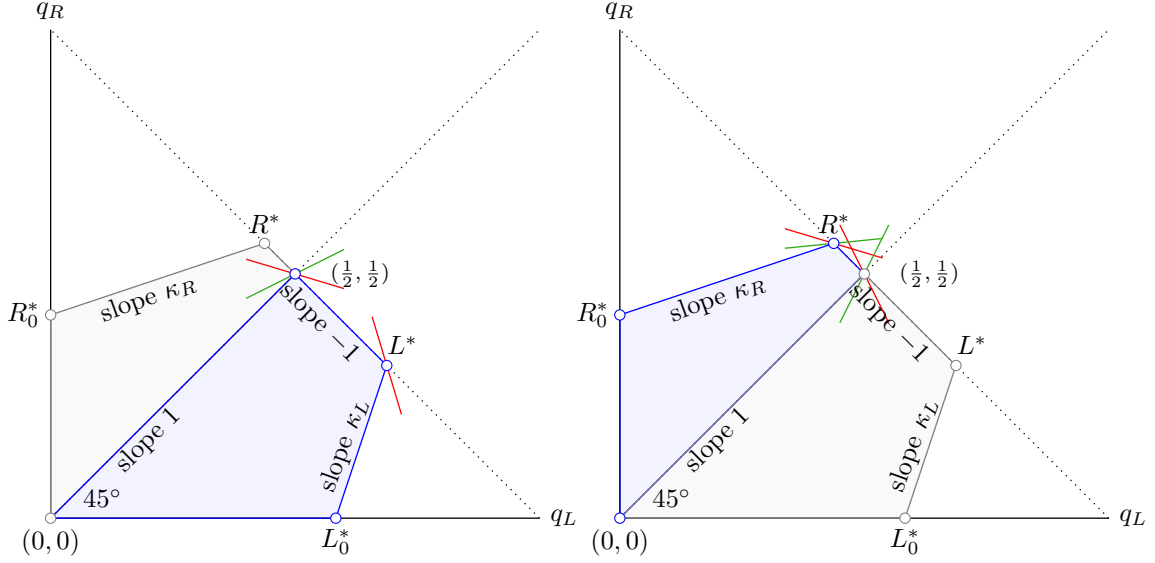

The most direct proof is graphical; throughout, we refer to the left panel of Figure \ref{fig:PropBLN_proof}. The feasible set $Q^-$ is shaded blue (the grey area is the part of the implementable set that is excluded by \eqref{eq:IR-theta_down}. Intermediary's objective function has slope $\Delta^-\equiv-(1-\phi^-(\theta))/\phi^-(\theta))$. Furthermore, at least one of $\{\phi^-(\theta),1-\phi^-(\theta)\}$ is positive, thus the objective function increases along at least one axis. 

\noindent\textit{Case 1: $\phi^-(\theta)\in[0,1]$}. If $\phi^-(\theta)\in[0,1]$, then $\phi^-(\theta)\geq 0$, $1-\phi^-(\theta)\geq 0$, and at least one inequality is strict. Thus, the objective function increases to the northeast. The optimum is at $(\half,\half)$ if $\Delta^->-1$ (the indifference curve is flatter than face (\ref{eq:feas1})) and it is $L^*$ if $\Delta^-<-1$ (indifference curve is steeper). If $\Delta^-=-1$, then $L^*$, $(\half,\half)$ or any convex combination of them is optimal (see red lines in the figure). Therefore,
\begin{itemize}
\item   If $\phi^-(\theta)\in[0,\half]$, then $(q_L^-(\theta),q_R^-(\theta))=L^*$.
\item If $\phi^-(\theta)\in[\half,1]$, then $(q_L^-(\theta),q_R^-(\theta))=(\half,\half)$.
\end{itemize}

\noindent\textit{Case 2: $\phi^-(\theta)\in[1,\phi^-(\thetah)]$}. If $\phi^-(\theta)\in[1,\phi^-(\thetah)]$, then $\phi^-(\theta)>0$ and $1-\phi^-(\theta)\leq 0$. Thus, the objective function increases to the northwest and $\Delta^-\in [0,1]$, so $(\half,\half)$ is optimal.

\noindent\textit{Case 3:} $\phi^-(\theta)\in[\phi^-(\thetal),0]$.  If $\phi^-(\theta)\in[\phi^-(\thetal),0]$, then $\phi^-(\theta)\leq 0$ and $1-\phi^-(\theta)> 0$. Thus, the objective function increases to the southeast and $\Delta^->0$. The optimum is at $L^*$ if $\Delta^->\kappa_L$ and is $L_0^*$ if $\Delta^-<\kappa_L$ (see green lines).  Thus, $L^*$ is optimal whenever,
\begin{align*}
\Delta^-<\kappa_L\Leftrightarrow -\frac{1-\phi^-(\theta)}{\phi^-(\theta)}<\kappa_L\Leftrightarrow 1-\phi^-(\theta)>-\kappa_L\phi^-(\theta)\Leftrightarrow 1>(1-\kappa_L)\phi^-(\theta).
\end{align*}
Note that if $\kappa_L<1$, then previous inequality obviously holds. Furthermore, if $\kappa_L>1$, the preceding inequality holds whenever
\begin{align*}
\phi^-(\theta)>\frac{1}{(1-\kappa_L)}=-\frac{1}{\kappa_L-1}.
\end{align*}
Because $\phi^-(\cdot)$ is increasing, this holds for all such $\phi^-(\theta)$ if and only if it holds at $\thetal$, which is guaranteed by (A\ref{a:bounds}) as $\phi^-(\thetal)=\thetal$. Thus, if $\phi^-(\theta)\in[\phi^-(\thetal),0]$, then $(q_L^-(\theta),q_R^-(\theta))=L^*$.

\noindent Combining Cases 1-3, the solution to the lower optimization is 
\begin{itemize}
\item   If $\phi^-(\theta)\leq \half$, then $(q_L^-(\theta),q_R^-(\theta))=L^*$.
\item If $\phi^-(\theta)\geq \half$, then $(q_L^-(\theta),q_R^-(\theta))=(\half,\half)$.
\end{itemize}

\noindent\textit{Step 2. We solve the upper optimization.} The upper optimization can be done pointwise, $\max_{(q_L(\theta),q_R(\theta)}u(q_L(\theta),q_R(\theta),\phi^+(\theta))$, subject to $(q_L(\theta),q_R(\theta))\in Q^+\equiv Q\cap \{q_L(\theta)\leq q_R(\theta)\}$.

The proof is again graphical, now using the right panel of Figure \ref{fig:PropBLN_proof}. The feasible set $Q^+$ is shaded blue (the grey area is the part of the implementable set that is excluded by (IR$\uparrow$). Intermediary's objective function has slope $\Delta^+\equiv-(1-\phi^+(\theta))/\phi^+(\theta))$. Furthermore, at least one of $\{\phi^+(\theta),1-\phi^+(\theta)\}$ is positive, thus the objective function increases along at least one axis. 

\noindent\textit{Case 1: $\phi^+(\theta)\in[0,1]$}. If $\phi^+(\theta)\in[0,1]$, then $\phi^+(\theta)\geq 0$, $1-\phi^+(\theta)\geq 0$, and one at least one inequality is strict. Thus, the objective function increases to the northeast and $\Delta^+0$. The optimum is at $R^*$ if $\Delta^+>-1$ (the indifference curve is flatter than face (\ref{eq:feas1})) and it is $(\half,\half)$ if $\Delta^+<-1$ (indifference curve is steeper). If $\Delta^+=-1$, then $R^*$, $(\half,\half)$ or any convex combination of them is optimal (see red lines in Figure \ref{fig:PropFB}). Therefore,
\begin{itemize}
\item   If $\phi^+(\theta)\in[0,\half]$, then $(q_L^+(\theta),q_R^+(\theta))=(\half,\half)$.
\item If $\phi^+(\theta)\in[\half,1]$, then $(q_L^+(\theta),q_R^+(\theta))=R^*$.
\end{itemize}

\noindent\textit{Case 2: $\phi^+(\theta)\in[1,\phi^+(\thetah)]$}. If $\phi^+(\theta)\in[1,\phi^+(\thetah)]$, then $\phi^+(\theta)> 0$, $1-\phi^+(\theta)\geq 0$. Thus, the objective function increases to the northwest and $\Delta^+\geq 0$. The optimum is at $R^*$ if $\Delta^+<\kappa_R$ and it is $R_0^*$ if $\Delta^+\geq\kappa_R$ (indifference curve is steeper). The solution is $R^*$ whenever, 
\begin{align*}
\Delta^+\leq \kappa_R\Leftrightarrow -\frac{1-\phi^+(\theta)}{\phi^+(\theta)}\leq \kappa_R\Leftrightarrow \phi^+(\theta)-1\leq\kappa_R\phi^+(\theta)\Leftrightarrow \phi^+(\theta)(1-\kappa_R)\leq 1.
\end{align*}
If $\kappa_R\geq 1$, then the preceding inequality is immediate. Suppose $\kappa_R<1$. Because $\phi^+(\cdot)$ is increasing, then the preceding holds for all such $\phi^+(\theta)$ if and only if it holds for $\phi^+(\thetah)$: 
\begin{align*}
\phi^+(\thetah)\leq \frac{1}{1-\kappa_R}\Leftrightarrow \thetah\leq \frac{1}{1-\kappa_R}.
\end{align*}
The preceding follows from (A\ref{a:bounds}). Thus,
 if $\phi^+(\theta)\in[1,\phi^+(\thetah)]$, then $(q_L^+(\theta),q_R^+(\theta))=R^*$.

\noindent\textit{Case 3: $\phi^+(\theta)\in[\phi^+(\thetal),0)$}.  If $\phi^+(\theta)\in[\phi^+(\thetal),0)$, then $\phi^+(\theta)\leq 0$ and $1-\phi^+(\theta)>0$. Thus, the objective function increases to the southeast and $\Delta^+>0$. The optimum is at $(\half,\half)$ if $\Delta^+>1$ and is $(0,0)$ if $\Delta^+<1$ (see green line in the figure).  Thus, $(\half,\half)$ is optimal whenever,
\begin{align*}
\Delta^+\geq 1\Leftrightarrow -\frac{1-\phi^+(\theta)}{\phi^+(\theta)}\geq 1\Leftrightarrow 1-\phi^+(\theta)\geq -\phi^+(\theta)\Leftrightarrow 1\geq 0.
\end{align*}
Note that the sign of the inequality is maintained in the second step because $-\phi^+(\theta)>0$. Hence, if $\phi^+(\theta)\in[\phi^+(\thetal),0)$, then $(q_L^+(\theta),q_R^+(\theta))=(\half,\half)$.

\noindent Combining Cases 1-3, the solution to the upper optimization is 
\begin{itemize}
\item   If $\phi^+(\theta)\leq \half$, then $(q_L^+(\theta),q_R^+(\theta))=(\half,\half)$.
\item If $\phi^+(\theta)\geq \half$, then $(q_L^+(\theta),q_R^+(\theta))=R^*$.
\end{itemize}

\noindent Recalling the definitions of $\tstar$ and $\tstarb$, we have the solution of the relaxed problem,
    \begin{align*}
(q_L(\theta),q_R(\theta))=\begin{cases}
  L^*  &\text{if}\qquad \theta\in[\thetal,\tstar)\\
  (\half,\half)  &\text{if}\qquad \theta\in[\tstar, \tstarb]\\
  R^* &\text{if}\qquad \theta\in(\tstarb,\thetah]
\end{cases}
\end{align*}
The construction of payments and $U(\cdot)$ is immediate, based on the discussion in the text.

\noindent\textit{Step 2. We verify that the solution of the relaxed problem satisfies (\ref{eq:MON})}. In the relaxed problem, for $\theta\in[\thetal,\tstar)$, the optimal influence bundle is $L^*$. In a \bln{} environment, $L^*$ lies below the 45\textdegree{} line. By implication for such types we have $q_R(\cdot)-q_L(\cdot)<0$ and constant. For $\theta\in(\tstar, \tstarb)$ the optimal influence bundle is $(\half,\half)$, and hence, $q_R(\cdot)-q_L(\cdot)=0$ for such types. For $\theta\in(\tstarb,\thetah)$, the optimal influence bundle is $R^*$. In a \bln{} environment, $R^*$ lies above the 45\textdegree{} line. By implication for such types we have $q_R(\cdot)-q_L(\cdot)>0$ and constant. (\ref{eq:MON}) follows.

\subsection{Proof of Proposition \ref{prop:unb}}
Having proved Proposition \ref{prop:bln}, we proceed more quickly. As described in Section \ref{sec:screening}, the intermediary's relaxed problem becomes
\begin{align}
\max_{(q_L(\cdot),q_R(\cdot))}\int_{\thetal}^{\thetah}u(q_L(\theta),q_R(\theta),\phi^{-}(\theta))f(\theta)\,d\theta\qquad\text{subject to \eqref{eq:I} and \eqref{eq:IR-theta_down}},
\end{align}
where $u(q_L(\theta),q_R(\theta),\phi^{-}(\theta))=\phi^{-}(\theta)q_R(\theta)+(1-\phi^{-}(\theta))q_L(\theta)$. 

We maximize this objective pointwise over choices $(q_L(\theta),q_R(\theta))\in Q\cap\{q_L(\theta)\geq q_R(\theta)\}$, which is the set of bundles satisfying \eqref{eq:I} and \eqref{eq:IR-theta_down}, shown in 
Figure \ref{Fig:Unb}.
Recall the definition $\Delta^-\equiv -(1-\phi^-(\theta))/\phi^-(\theta)$ from the proof of Proposition \ref{prop:bln} as the slope of the objective function's indifference curves. Consider the three intervals of types in the proposed mechanism.

For $\theta<\tstar$, we have $\phi^-(\theta)<\frac{1}{2}$. We argue that $L^*$ is the maximizer. If $\phi^-(\theta)\leq 0$, then the objective is increasing in $q_L$ and decreasing in $q_R$, but since $\phi^-(\theta)>\theta\geq \underline\theta$, the slope is greater than $\kappa_L$ by Assumption \ref{a:bounds}. Thus, if $\phi^-(\theta)\leq 0$, $L^*$ is the maximizer. If instead $\phi^-(\theta)\in (0,\frac{1}{2})$, then the objective is increasing in both dimensions with slope $\Delta^-<-1$, and again the maximizer is $L^*$.

For $\theta \in (\tstar,\tstaru)$, we have $\frac{1}{2}<\phi^-(\theta)<\frac{1}{1-\kappa_R}$. Hence, the objective is increasing in at least $q_R$, and the indifference curves are either decreasing with slope $\Delta^->-1$ (when $\phi^-(\theta)<1$ or are increasing with slope $\Delta^-<\kappa_R$ (when $\phi^-(\theta)>1$). In either case, it is clear by inspection of Figure \ref{Fig:Unb} that $R^*$ is the maximizer.

For $\theta\in (\tstaru,\overline\theta]$, assuming this interval is nonempty, we have $\phi^-(\theta)>\frac{1}{1-\kappa_R}>1$. Thus, the objective is increasing in $q_R$ and decreasing in $q_L$, with slope $\Delta^->\kappa_R$. By inspection of the same figure, $B$ is the maximizer.

By construction, this set of allocations $\{(q_L(\theta),q_R(\theta))\}_{\theta\in\Theta}$ solves the relaxed problem. To verify that it solves the original problem, we need only verify that \eqref{eq:MON} is satisfied. Note that $q_R(\cdot)-q_L(\cdot)$ is constant on each of the three intervals $[\thetal,\tstar)$, $(\tstar,\tstaru)$, and $(\tstaru,\thetah]$. On the first two intervals, we have $q_L(\cdot)=1-q_R(\cdot)$, so it suffices to show that $q_R(\cdot)$ is higher under $R^*$ than under $L^*$, but this is immediate, since $R^*$ maximizes $q_R$ over the set $Q$ (which contains $L^*$). Thus, $q_R(\cdot)-q_L(\cdot)$ has the correct monotonicity on the first two intervals. This property extends to the third interval; indeed, in an \unb{} environment, $R^*$ lies below the 45\textdegree{} line $(q_R(\cdot)-q_L(\cdot)<0)$ while $B$ lies on it $(q_R(\cdot)-q_L(\cdot)=0)$.

\bibliographystyle{aer} 
\bibliography{intermediary}
\end{document}